\documentclass[aps,prb,preprint,preprintnumbers,superscriptaddress,amsmath,amssymb,showpacs]{revtex4}%

\newcommand{\LL}{\mathcal{L}}
\newcommand{\PP}{\mathcal{P}}

\newcommand{\dd}{\mathrm{d}}

\newcommand{\ii}{\text{i}}

\newcommand{\tr}[2]{\text{Tr}_{ #1 } \left\{ #2 \right\}}

\newcommand{\av}[1]{\langle #1 \rangle}

\newcommand{\expo}[1]{\text{exp}\left( #1 \right)}
\newcommand{\stat}{\text{st}}
\newcommand{\tS}{\text{S}}
\newcommand{\tL}{\text{L}}
\newcommand{\tR}{\text{R}}

\newcommand{\tb}{\text{B}}
\newcommand{\tsb}{\text{SB}}
\newcommand{\sta}{\text{st}}

\newcommand{\id}{\mathds{1}}
\newcommand{\e}{\text{e}}

\newcommand{\unit}[1]{\,\mathrm{#1}}

\usepackage{braket}
\usepackage{subfigure}
\usepackage{graphicx}
\usepackage{xcolor}
\usepackage{dsfont}
\usepackage{pstricks}

\graphicspath{{./Plots_FC/}{./Plots_1m/}{./Plots_mm/}{./Plots_lo/}{./Plots_mm_F_V/}{./Plots_1m_fig/}{./Plots_BDBT/}}

\begin{document}


\title{Current Noise in Single-Molecule Junctions Induced by Electronic-Vibrational Coupling}

\author{C.\ Schinabeck}
\affiliation{Institut f\"ur Theoretische Physik und Interdisziplin\"ares Zentrum f\"ur Molekulare
Materialien, \\
Friedrich-Alexander-Universit\"at Erlangen-N\"urnberg,\\
Staudtstr.\, 7/B2, D-91058 Erlangen, Germany
}
\author{R.\ H\"artle}
\altaffiliation[Current address: ]{Institut f\"ur Theoretische Physik, Friedrich-Hund-Platz 1, D-37077 G\"ottingen, Germany}
\affiliation{Institut f\"ur Theoretische Physik und Interdisziplin\"ares Zentrum f\"ur Molekulare
Materialien, \\
Friedrich-Alexander-Universit\"at Erlangen-N\"urnberg,\\
Staudtstr.\, 7/B2, D-91058 Erlangen, Germany
}

\author{H.\ B.\ Weber}
\affiliation{Lehrstuhl f\"ur Angewandte Physik,\\ Friedrich-Alexander-Universit\"at Erlangen-N\"urnberg,\\ Staudtstr.\, 7/A3, D-91058 Erlangen, Germany}
\author{M.\ Thoss}
\affiliation{Institut f\"ur Theoretische Physik und Interdisziplin\"ares Zentrum f\"ur Molekulare
Materialien, \\
Friedrich-Alexander-Universit\"at Erlangen-N\"urnberg,\\
Staudtstr.\, 7/B2, D-91058 Erlangen, Germany
}

\author{}
\affiliation{}


\date{\today}

\begin{abstract}
The influence of multiple vibrational modes on current fluctuations in electron transport through single-molecule junctions is investigated. 
Our analysis is based on a generic model of a molecular junction, which comprises a single electronic state on the molecular bridge coupled to multiple vibrational modes and fermionic leads, and employs a master equation approach. The results reveal that in molecular junctions with multiple vibrational modes already
weak to moderate electronic-vibrational coupling may result in high noise levels, especially at the onset of resonant transport, in accordance with experimental findings of Secker \emph{et al.}.\cite{Secker2011}
The underlying mechanisms are analyzed in some detail.
\end{abstract}


\maketitle

\section{Introduction}

Any measurement is accompanied by temporal fluctuations of the recorded signal. These fluctuations may be a valuable source of information far beyond the time averaged value of the signal.\cite{Landauer1998,Blanter2000,Beenakker2003,cuevasscheer2010,Nazarov2009}
For example, measurements of current noise  allowed to determine the value of the elementary charge \cite{Hartmann1921,Hull1925} and the charge of a Cooper pair \cite{Lefloch2003} as well as to verify the existence of other quasiparticles, such as, for example, in quantum Hall systems.\cite{Laughlin1983,Saminadayar1997,Picciotto1997} Noise measurements in nanostructures facilitated the determination of the number and transparency of transmission channels e.g. in atomic-sized metal contacts,\cite{Brom1999,Wheeler2010,Schneider2010,Chen2012} mesoscopic diffusive wires \cite{Beenakker1992,Henny1999,Liefrink1994,Nagaev1992,Steinbach1996,Cron2001} as well as single-molecule junctions, i.e.\ systems where single moelcules are chemically bound to metal or semiconductor electrodes.\cite{Djukic2006,Kiguchi2008,Tal2008}  Single molecule junctions, which have been experimentally realized by various techniques,\cite{Reed1997,Reichert2002,Natelson04,Sapmaz2005,Ho98,Wu04,Mayor2009,Chen2009} will be the focus of this paper.

A particularly interesting aspect of charge transport in molecular junctions is the crucial importance of electron-vibrational coupling.
Due to their small size and mass, the transport characteristics of single molecule junctions are usually characterized by an intricate interplay between electronic and vibrational degrees of freedom.\cite{Galperin05,Galperin07,Leon08,Ioffe08,Natelson08,Haertle09,Huettel09,Repp10,Tao10,Ballmann10,Osorio10,Arroyo10,Secker2011,Haertle2011,Haertle2011c,Kim2011,Ward11,Ivan2011,Ballmann2012,Albrecht12,Wilner2013,Ballmann13,Haertle2013,Haertle13b}
Vibrational signatures have also been reported in noise measurements in molecular junctions.\cite{Tal2008,Kumar2012,Sperl2010,Tsutsui2010,Neel2011} 
Morevover, it has been shown that molecular junctions may exhibit very large noise levels, \cite{Secker2011,Sydoruk2012} which exceed the Poissonian limit (corresponding to statistically uncorrelated tunneling events) by several orders of magnitude. Theoretical studies have shown that such high noise levels in molecular junctions can be caused by
electronic-vibrational coupling.\cite{Koch2005,Koch2006,Donarini2005,Brueggemann2012,Schaller2013}
However, it has also been shown, that electronic-vibrational coupling may results in the opposite effect, i.e.\ an anomalous suppression of the current noise below the Poissonian limit due to the interplay of vibrational degrees of freedom and relaxation.\cite{Haupt2006,Haupt2008}

In this paper, we investigate current fluctuations in vibrationally coupled electron transport through single-molecule junctions and adress the physical origin of the large noise levels reported by Secker \emph{et al.}.\cite{Secker2011} Extending previous work,\cite{Koch2005,Koch2006,Donarini2005,Brueggemann2012,Schaller2013} which was limited to a single vibrational mode, we specifically investigate the effect of multiple vibrational modes. Our studies are inspired by the work of Koch \emph{et al.},\cite{Koch2005,Koch2006}
which predicted the occurence of high noise levels in model molecular junctions where a single electronic level
is strongly coupled to a single vibrational mode. The electronic-vibrational coupling strength required to obtain
these high noise levels was, however, very large. Furthermore, molecular junctions typically involve a multitude of vibrational modes. We, therefore, explore the possibility that electronic-vibrational coupling to multiple vibrational modes can result in high noise levels even for small to moderate electron-vibrational coupling. 

In our study, we employ a master equation approach developed by Flindt \emph{et al.}\ \cite{Flindt2005E,Flindt2004}
This approach is based on the MacDonald formula,\cite{MacDonald1949} which relates the noise power to the moments of the distribution of the transferred charge, the so-called full counting statistics (FCS) \cite{Levitov1996,Bagrets2003,Belzig2005,Emary2007,Welack2008,Haupt2009,Avriller2009,FlindtNovotny2009,Marcos2010,Flindt2010,Avriller2010,Urban2010,Park2011,Emary2011} of electron transport through a nanoscale electronic device. Thereby, the electronic-vibrational coupling is treated non-perturbatively, whereas a second-order expansion in the molecule-lead coupling is applied, thus limiting the study to resonant transport processes.
It is noted that a variety of other theoretical approaches exist to investigate current fluctuations
of a nanoelectronic conductor. This includes other master equation approaches,\cite{Hershfield1993,Korotkov1994,Haupt2006} scattering theory \cite{Blanter2000, Buettiker1992} and nonequilibrium Green's function theory.\cite{Meir1992,Runge1993,GalperinShot2006,Haug2008,Novotny2011}

The paper is organized as follows. In Sec.\ \ref{sec:theory_H}, we introduce the model Hamiltonian used to describe electron transport through single-molecule junctions. The  master equation approach and its application to the calculation of the observables of interest, especially mean current and noise is outlined in Secs.\ \ref{sec:theory_MEQ} and \ref{sec:theory_OoI}. The results are presented in Sec.\ \ref{sec:results}.
Thereby, we first review basic concepts of noise (Sec.\ \ref{sec:basic_concepts}) and provide a detailed analysis of the effects for a single vibrational mode (Sec.\ \ref{sec:results_1m} ). 
The influence of multiple vibrational modes on the current noise is investigated in Sec.\ \ref{sec:results_mm}.
Throughout the paper, we use units where $\hbar=1$ and $e=1$.

\section{Theoretical methodology} \label{sec:theory}

\subsection{Model Hamiltonian} \label{sec:theory_H}
In order to investigate vibrationally coupled electron transport in molecular junctions, we employ the following model Hamiltonian for a molecular junction:\cite{Braig2003,Mitra2004,Cizek2004,Koch2005,Wegewijs2005,Haertle2008}
\begin{align}
 H=\epsilon_0 d^{\dagger} d +\sum_{\alpha=1}^M \Omega_\alpha a^{\dagger}_\alpha a_\alpha +\sum_{\alpha=1}^M \lambda_\alpha (a_\alpha +a_\alpha^\dagger) d^{\dagger} d +\sum_{k \in L/R} \epsilon_k c_k^{\dagger} c_k +\sum_{k \in L/R} (V_k c_k^{\dagger} d +V_k^\ast d^{\dagger} c_k).
\end{align}
Thereby, a single electronic level with energy $\epsilon_0$ located on the molecular bridge is coupled to a continuum of electronic states in the macroscopic leads via interaction matrix elements $V_k$. The energy of these lead states is given by $\epsilon_k$. The operators $d^\dagger/d$ and $c_k^\dagger/c_k$ denote the creation / annihilation operators for the single electronic state on the molecular bridge and the states in the leads, respectively. The vibrational degrees of freedom of the molecular junction are described by $M$ harmonic modes with frequencies $\Omega_\alpha$ and corresponding creation and annihilation operators $a_\alpha^\dagger/a_\alpha$. The coupling strengths between the harmonic modes and the single electronic state are given by $\lambda_\alpha$. The
inclusion of only a single electronic level on the molecular bridge  is supported by experimental results of Secker \emph{et al.}, which report the highest noise levels before the first electronic state enters the bias window in the experiment.\cite{Secker2011}

For vanishing molecule-lead coupling, $V_k \to 0$, the Hamiltonian $H$ can be diagonalized analytically by the small polaron transformation, 
\begin{eqnarray}
H \to \bar H = U H U^\dagger=\bar H_\tS +\bar H_\tb+ \bar H_\tsb
\end{eqnarray} 
with
\begin{subequations}
 \begin{align}
   U&=\expo{d^\dagger d \sum_{\alpha=1}^{M} \frac{\lambda_\alpha}{\Omega_\alpha} \left(a_\alpha^\dagger -  a_\alpha \right)},\\
  \bar H_\tS&=\bar \epsilon_0 d^\dagger d + \sum_{\alpha=1}^M \Omega_\alpha a^{\dagger}_\alpha a_\alpha,\\
  \bar H_\tb&=\sum_{k \in L/R} \epsilon_k c_k^{\dagger} c_k,\\
  \bar H_\tsb&=\sum_k (V_{k} X c_k^\dagger d + \text{h.c.}).
 \end{align}
\end{subequations}
Thereby, we have partitioned the Hamiltonian into three parts: The system part $\bar H_\tS$ comprises the electronic state at the molecular bridge with the polaron shifted energy $\bar \epsilon_0=\epsilon_0-\sum_\alpha \frac{\lambda_\alpha^2}{\Omega_\alpha}$ and the $M$ vibrational modes. The term $\bar H_\tb$ refers to the degrees of freedom of the leads. The coupling between the leads and the molecule is described by $\bar H_\tsb$, which is renormalized by the overall shift operator $X=\prod_\alpha X_\alpha$ with $X_\alpha=\exp\{(\lambda_\alpha^2/\Omega_\alpha)(a_\alpha-a_\alpha^\dagger)\}$. The interaction between the molecule and the left and the right leads, respectively, is characterized by the level width functions $\Gamma_{\tL/\tR} (E)=2 \pi \sum_{k \in \tL/\tR} | V_k |^2 \delta (E-\epsilon_k)$. For the remainder of the paper, the wide-band approximation is applied, which implies constant level width functions $\Gamma_{\tL/\tR}$. Assuming, furthermore,  that the molecule is coupled symmetrically to the leads, a value of $\Gamma_{\tL/\tR}=0.2 \unit{meV}$ is used. As most experiments on single-molecule junctions are performed at low temperatures, the temperature of the leads is set to $10 \unit{K}$. Consequently, $k_\tb T\gg \Gamma \equiv \Gamma_\tL+\Gamma_\tR$ always holds, which is a prerequisite for the validity of the second-order master equation approach, which is introduced in the following section. This also means that the temperature induced broadening exceeds the energy level broadening due to the coupling to the leads (described by $\Gamma$) at the onset of the resonant transport regime.
Additionally, we assume a symmetric drop of the bias voltage at the contacts.
%

\subsection{Master Equation Formalism} \label{sec:theory_MEQ}

Based on the model introduced above, we use a master equation (ME) approach to study transport processes. Within this approach, the observables of interest, especially the average current and the zero-frequency noise,\cite{Hershfield1993,Korotkov1994,Flindt2004,Flindt2005E} are evaluated using the reduced density matrix $\rho$, which is obtained by taking the trace of the total density matrix of the overall system over the electronic degrees of freedom of the leads. This involves the
calculation of expectation values $\av{A}$ and two-time correlation functions $\av{A(t') B(t)}$, where $A$ and $B$ are operators of the overall system. As these quantities are invariant under unitary transformations like the small Polaron transformation, the transformed Hamiltonian $\bar H$ can be considered as the Hamiltonian of the overall system and thus the reduced density matrix is determined by the following equation of motion,\cite{Lehmann2004,Mitra2004,Haertle2009,May2002,Harbola2006,Volkovich2008,Haertle2010,Haertle2011c,Egorova2003}
\begin{equation}
\frac{\partial \rho (t)}{\partial t}= -i [\bar H_\tS, \rho(t)]_- - \int_0^\infty \dd \tau \, \text{Tr}_\tb \{ [\bar H_{\tsb},[\bar H_{\tsb}(-\tau), \rho(t) \rho_\tb]_- ]_- \} \equiv \LL \rho (t)
\label{eq:ME}
\end{equation}
with
\begin{equation}
\bar H_\tsb (-\tau) =\e^{-\ii \tau (\bar H_\tS +\bar H_\tb)} \bar H_\tsb \e^{\ii \tau (\bar H_\tS +\bar H_\tb)}.
\end{equation}
Here, $\rho_\tb$ denotes the equilibrium density matrix of the leads, given by
\begin{equation}
 \rho_\tb = Z^{-1} \e^{-( \bar H_\tb - \mu_\tL N_\tL - \mu_\tR N_\tR)/ (k_B T)}, \quad Z=\tr{\tb}{\e^{-( \bar H_\tb - \mu_\tL N_\tL - \mu_\tR N_\tR)/ (k_B T)}},
\end{equation}
where $N_{\tL / \tR}=\sum_{k \in \tL / \tR} c_k^\dagger c_k$ represents the occupation number operator of the left and the right lead, respectively.
Furthermore, we have defined the Liouvillian $\LL$. Eq.\ (\ref{eq:ME}) can be derived based on the well-known Nakajima-Zwanzig equation,\cite{Nakajima1958,Zwanzig1960} employing a second-order expansion in the coupling $\bar H_\tsb$ along with the so-called Markov approximation.
Due to expansion to second order in the molecule-lead coupling, the ME (\ref{eq:ME}) only comprises resonant transport processes whereas it misses higher order processes like cotunneling which especially dominate in the nonresonant transport regime.
In this paper, we focus on the steady state, $\rho^\stat \equiv \rho (t \to \infty)$, which is reached at long times and is obtained by solving Eq. \  (\ref{eq:ME}) with the left hand side set to zero, $\partial \rho (t \to \infty) / \partial t=0$.
Details on the explicit evaluation of the ME in the product basis of the electronic and vibrational degrees of freedom can be found in Ref.\ \onlinecite{Haertle2011}.

In the following, we neglect vibrational coherences, i.e. the vibrational off-diagonal elements of the density matrix. 
This is justified as long as the level width function $\Gamma=\Gamma_\tL+\Gamma_\tR$ is the smallest energy scale of the system and thus the relation $\Gamma \ll \Omega_\alpha$ holds. In the present case, this is no additional restriction as it is already a requirement for the validity of the second-order ME.\cite{Egorova2003} We will also consider the coupling to a multitude of vibrational modes. In this case, we have chosen the frequencies of the involved modes in such a way that the low integer multiples of different frequencies do not coincide. In that way, we avoid the quasi-degeneracy of excited levels so that the relation $\Gamma \ll |n \Omega_\alpha -m\Omega_{\alpha'}|$ holds for small integers $m$ and $n$ ($\alpha \neq \alpha'$). As a consequence, coherences are negligible in these systems, as was shown by H\"artle \emph{et al.}.\cite{Haertle2011}
\subsection{Full Counting Statistics and Observables of Interest} \label{sec:theory_OoI}

In the long time limit, the full information about electron transport is encoded in the distribution of electrons which have tunneled across the molecular bridge, the so-called full counting statistics (FCS). To obtain this distribution, $P_\tR (t,n)$, we ``count'' the number $n$ of electrons which are collected in the right lead during the time span $[0,t]$.
In line with the FCS the ME $\dot \rho(t)=\LL \rho(t)$ can be resolved with respect to this number $n$ of electrons ($n$-resolved ME).
To this end, the Liouvillian can be decomposed into three parts as described in Refs.\ \onlinecite{Marcos2010,Flindt2005}
\begin{equation}
 \LL=\LL_0 +\mathcal{I}^+_\tR + \mathcal{I}^-_\tR,
\label{eq:res_L}
\end{equation}
where the particle current superoperators $\mathcal{I}_\tR^\pm$ refer to physical processes in which one electron is created (+) or annihilated (-) in the right lead and $\LL_0$ corresponds to the part in which the number of electrons is not changed. Consequently, the $n$-resolved ME can be written as
\begin{align}
 \dot \rho_n(t)&=\LL_0 \rho_n(t) + \mathcal{I}^+_\tR \rho_{n-1}(t) + \mathcal{I}^-_\tR \rho_{n+1}(t) \nonumber \\
&=(\LL-\mathcal{I}^+_\tR -\mathcal{I}^-_\tR)\rho_n(t) + \mathcal{I}^+_\tR \rho_{n-1}(t) + \mathcal{I}^-_\tR \rho_{n+1}(t), \label{eq:nMEQ}
\end{align}
where the $n$-resolved density matrix $\rho_n$ fulfills the sum rule $\sum_n \rho_n(t)=\rho (t)$.
%
%
Applying the trace over the system degrees of freedom in Eq.\ (\ref{eq:nMEQ}), an EOM for the probability distribution $P_\tR(t,n)=\tr{\tS}{\rho_n(t)}$ is obtained
\begin{equation}
 \dot P_\tR(t,n)=\tr{\tS}{ \mathcal{I}^+_\tR [\rho_{n-1}(t) -\rho_n (t)]+\mathcal{I}^-_\tR [\rho_{n+1}(t) -\rho_n (t)]},
\label{eq:P_n}
\end{equation}
where the relation $\tr{\tS}{\LL \rho_n(t)}=0$ has been used.\cite{Flindt2004,Flindt2010}

Within this FCS-approach the average total current $\av{I}$ across the molecular bridge, which is independent of time in the stationary state, is given by
\begin{equation}
 \av{I}=\av{I_\tR}=\av{I_\tL}=\frac{\dd}{\dd t} \av{Q_\tR (t)},
\label{eq:meancurrent}
\end{equation}
where $\av{I_\tL}$ and $\av{I_\tR}$ are the average particle currents across the left and the right junction, respectively. Thereby, $\av{Q_\tR (t)}$ denotes the mean amount of charge collected in the right lead within the time span $[0,t]$ and is thus identical with the first moment ($\alpha=1$) of the distribution $P_\tR (t,n)$
\begin{equation}
 \av{Q_\tR^\alpha (t)}=\sum_n n^\alpha P_{\tR} (t,n). 
\end{equation}
In order to evaluate Eq.\ (\ref{eq:meancurrent}), we substitute the EOM for the probability distribution $P_\tR(t,n)$ from Eq.\ (\ref{eq:P_n}) together with $\rho^\stat=\sum_n \rho_n (t)$ and thus obtain 

\begin{align}
 \av{I}=\frac{\dd}{\dd t} \av{Q_\tR (t)}&=\sum_n n \dot P_\tR(t,n) \nonumber\\
&=\tr{\tS}{ \left(\mathcal{I}_\tR^+ - \mathcal{I}_\tR^- \right)  \sum_n \rho_n(t)}=\tr{\tS}{\left(\mathcal{I}_\tR^+ - \mathcal{I}_\tR^- \right) \rho^\sta}. \label{eq:expression1}
\end{align}
%

In the stationary limit the noise power spectral density $S_I (\omega)$ of the total current $I$ is given by the Fourier transform of the symmetrized current autocorrelation function
according to the Wiener-Khintchine theorem \cite{Wiener1930,Khintchine1934,Brown1992}
\begin{equation}
  S_I (\omega)=\int_{-\infty} ^\infty \dd \tau \frac{1}{2} \langle\left[ \Delta I(\tau), \Delta I(0) \right]_+\rangle \e^{\ii \omega \tau}.
\label{eq:PSD}
\end{equation}
Thereby, $\Delta I(t)=I(t) - \av{I}$ denotes the time-dependent current fluctuation operator (zero-mean current) in the Heisenberg picture.\\
As the zero-frequency power spectral density refers to the long time limit $t \to \infty$ (cf.\ Eq.\ (\ref{eq:MDF_S0})), the particle currents are conserved and equal to the total current $I(t)|_{t \to \infty}=I_\tL(t)|_{t \to \infty}=I_\tR(t)|_{t \to \infty}$ so that $S_I(0)=S_{I_\tL}(0)=S_{I_\tR} (0)$ holds.\cite{Blanter2000} 
Applying the MacDonald formula \cite{MacDonald1949,Flindt2005E} (cf.\ Eq.\ (\ref{eq:app_MDF_S}) in the appendix), which relates the noise power spectral density $S_{I_\tR} (\omega)$ to the moments $\av{Q_\tR^\alpha (t)}$ of the FCS, the following expression for the zero-frequency noise is obtained \cite{Flindt2005E}
\begin{align}
S_{I_\tR} (0)&=\frac{\dd}{\dd t} \left[\langle Q_\tR^2 (t)\rangle -\langle Q_\tR (t) \rangle ^2 \right] |_{t \to \infty}. \label{eq:MDF_S0}
\end{align}
Following the derivation of Flindt \emph{et al.},\cite{Flindt2004, Flindt2005E} the power spectral density of the zero-frequency noise can be expressed as
\begin{align}
 S_I (\omega=0)=S_{I_\tR} (0)
&=\tr{\tS}{\left( \mathcal{I}_\tR^+ +\mathcal{I}_\tR^- \right)\rho^\stat} -2 \tr{\tS}{ \left( \mathcal{I}_\tR^+ -\mathcal{I}_\tR^- \right) \mathcal{R} \left( \mathcal{I}_\tR^+ -\mathcal{I}_\tR^- \right) \rho^\stat}, \label{eq:S_0}
\end{align}
where $\mathcal{R}(0)=(1-\PP) \LL^{-1} (1-\PP)$ denotes the pseudoinverse of $\LL$.
As the orthogonal projection operator $\PP$ is defined as \cite{Flindt2004,Flindt2005E}
\begin{equation}
  \PP=\ket{0}\rangle\langle\bra{\tilde 0},
\label{eq:P}
\end{equation}
the operators $1-\PP$ ensure that the inversion only takes place within the subspace where the Liouvillian $\LL$ is not singular. The doublebrackets in Eq.\ (\ref{eq:P}) distinguish vectors in Liouville space from ordinary vectors in Hilbert space. Thereby, $\ket{0} \rangle \equiv \rho^\stat$ and $\langle \bra{\tilde 0}=\id$ denote the left and the right eigenvector of $\LL$ with eigenvalue zero in terms of the Liouville space. These eigenvectors obey the normalization constraint $\langle \braket{\tilde 0 | 0} \rangle=1$ and they are not neccessarily Hermitian conjugates of each other because $\LL$ is non-Hermitian in general.
The pseudoinverse $\mathcal{R}$ is identical with the well-known Moore-Penrose pseudoinverse $\mathcal{R}_{\text{MP}}$ of $\LL$ projected onto the regular subspace, i.e. $\mathcal{R}=(1-\PP) \mathcal{R}_{\text{MP}} (1-\PP)$.\cite{Flindt2010}
The details of the derivation \cite{Flindt2004,Flindt2005E} are summarized in the appendix for convenience.

To analyze the noise properties, it is useful to introduce the Fano factor $F$, which is given as the ratio between the actual zero-frequency noise $S_I(\omega=0)$ and the ``white'' noise of a Poisson process $S_I^\text{ps}=S_I^\text{ps}(\omega)=\av{I}$
\begin{equation}
 F=\frac{S_\text{I}(\omega=0)}{S_I^\text{ps}}=\frac{S_\text{I}(\omega=0)}{\av{I}}.
\end{equation}
\section{Results} \label{sec:results}
In this section we use the methodology introduced in Sec.\ \ref{sec:theory} to investigate the transport properties of single-molecule junctions. Thereby, we mainly focus on the zero-frequency noise expressed by the corresponding Fano factor and the current-voltage characteristics. As single-molecule junctions typically comprise a multitude of vibrational modes, we focus on the investigation of effects which result from the coupling to multiple vibrational modes (``multimode effects''). This is the topic of Sec.\ \ref{sec:results_mm}. To facilitate the discussion of multimode effects, we first consider two simpler systems. In Sec.\ \ref{sec:basic_concepts}, we introduce the basic concepts of noise with purely electronic transport and in Sec.\ \ref{sec:results_1m} we consider the case of strong ($\frac{\lambda}{\Omega}=4$) electronic-vibrational coupling to a single vibrational mode in order to introduce the phenomenon of avalanche-like transport \cite{Koch2005} and the basic transport mechanisms in this regime.

%
%
\begin{figure}[htbp]
  \centering
\begin{tabular}{l}
 (a)\\
     \includegraphics[width=.7\textwidth]{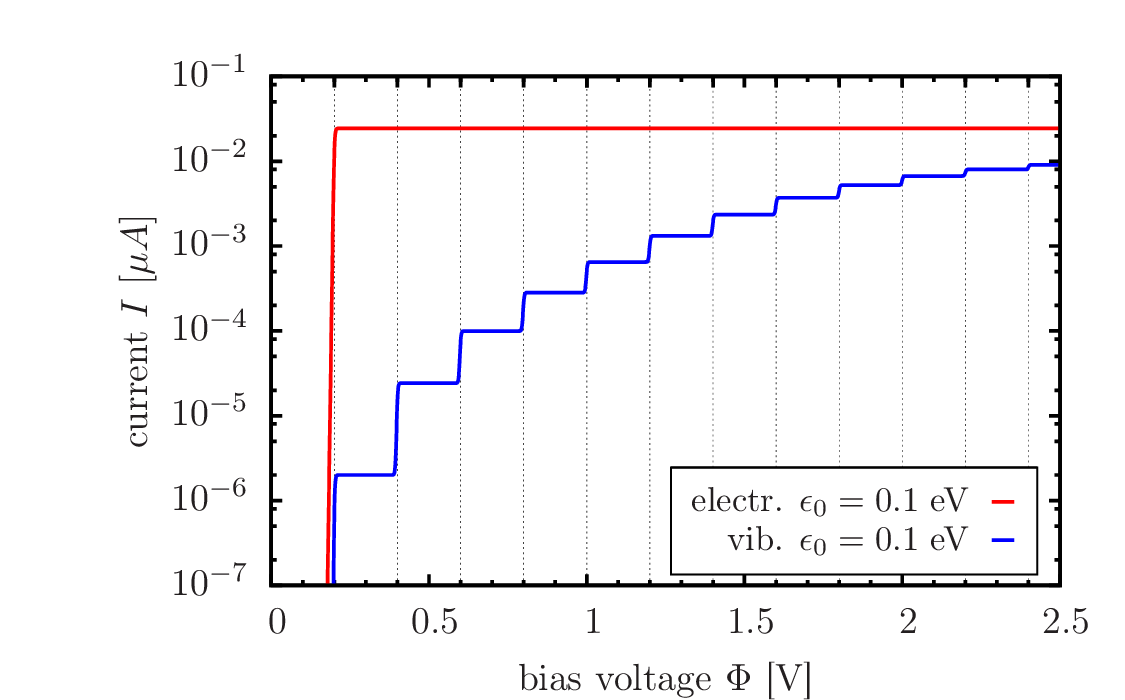} 
\\ 
 (b)\\
    \includegraphics[width=.7\textwidth]{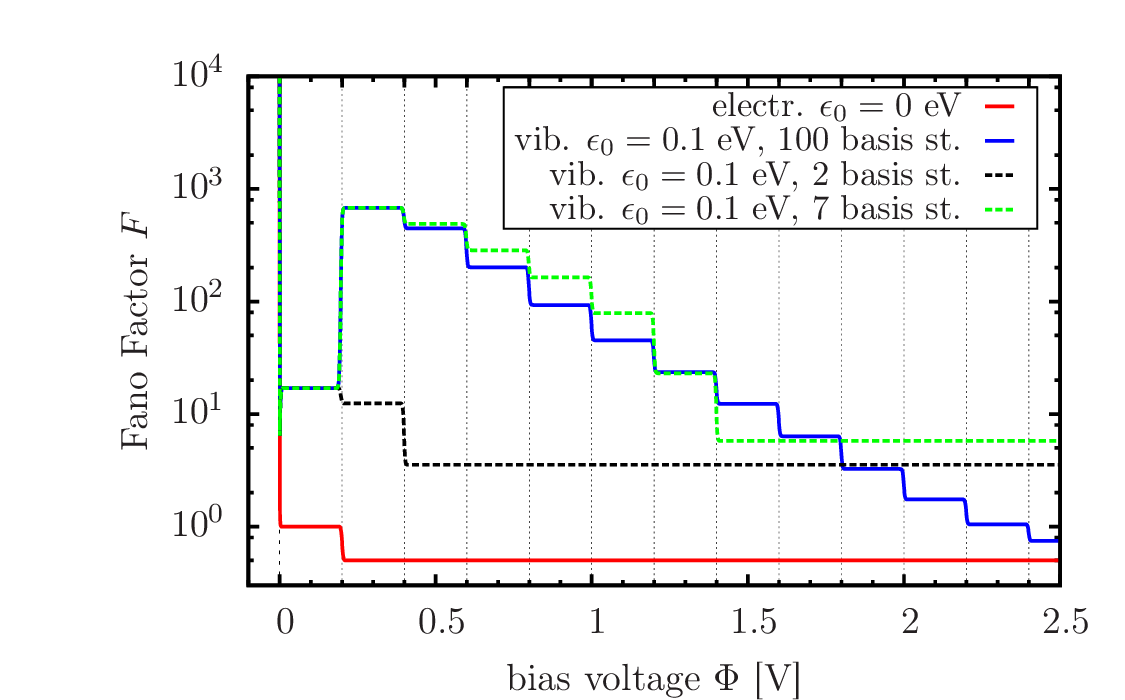}
\end{tabular}
  \caption{Current-voltage-characteristics (a) and Fano factor-voltage characteristics (b), calculated for purely electronic transport (electr.) as well as for the strong coupling ($\lambda/\Omega=4$) to a single vibrational mode with frequency $\Omega=0.1\unit{eV}$ (vib.). In the latter case, the dashed lines refer to a restriction of the vibrational basis set to 2 and 7 states, respectively.}
  \label{fig:4lo_I_F_el}
\end{figure}

\subsection{Basic Concepts of Noise for a Purely Electronic System} \label{sec:basic_concepts}

In order to introduce the basic concepts, we first discuss purely electronic transport (i.e.\ without electronic-vibrational coupling) through a single electronic state with energy $\bar \epsilon_0=0.1 \unit{eV}$. The red line in Fig.\ \ref{fig:4lo_I_F_el}a presents the corresponding current-voltage characteristics. For bias voltages $\Phi < 2 \bar \epsilon_0$ the current is almost zero because the electronic level is outside the bias window, which is given by the chemical potentials $\mu_{\tL/\tR}=\pm \Phi/2$. This is the nonresonant transport regime because resonant transport processes can only occur due to the thermal broadening of the Fermi distribution in the left lead and are thus very rare. Higher-order nonresonant transport processes such as cotunneling, which may dominate transport in this regime for lower temperatures, are not included in our second 
order ME approach.
At $\Phi=2 \bar \epsilon_0$ the electronic level enters the bias window so that the current increases in one step to a constant value. This indicates the onset of the resonant transport regime ($\Phi> 2 \bar \epsilon_0$). 

Additional information about the transport mechanisms in these two regimes can be obtained from the Fano factor-voltage characteristics, which is shown in Fig.\ \ref{fig:4lo_I_F_el}b. At bias voltage $\Phi=0$ the Fano factor-voltage characteristics shows a singularity corresponding to the finite thermal Nyquist-Johnson noise \cite{Blanter2000,cuevasscheer2010} which is due to the thermal fluctuations of the occupation number in the left and the right lead, whereas for higher bias voltages the nonequilibrium or shot noise contribution dominates.
In the nonresonant transport regime, the Fano factor assumes a constant value of unity,
which indicates that electron transport obeys Poissonian statistics and is thus statistically uncorrelated.\cite{Blanter2000,cuevasscheer2010}
At the cross-over between the nonresonant and the resonant regime the Fano factor decreases in a single step to a constant value of 1/2. This limit corresponds to the regime of the Pauli blockade where electron transport is anti-correlated due to the Pauli exclusion principle.\cite{cuevasscheer2010}


%
\subsection{Electronic-Vibrational Coupling to a Single Mode} \label{sec:results_1m}

In the following we extend our model system and consider coupling of the electronic state to a single vibrational mode with frequency $\Omega=0.1 \unit{eV}$. The regime of moderate electronic-vibrational coupling ($\lambda / \Omega \leq 1$) has been discussed elsewhere.\cite{Haertle2011,Mitra2004,Park2011} In order to analyze the regime of avalanche-like transport, we focus on the strong dimensionless coupling $\lambda/\Omega=4$. To this end, Fano factor-voltage characteristics are analyzed in Sec.\ \ref{subsubsec:1m_res} for the resonant transport regime. Thereby, we review the results obtained by Koch \emph{et al.} \cite{Koch2005,Koch2005FCS,Koch2006,Koch2006PRL} and analyze the structure of avalanche-like transport from a somewhat different perspective. This is followed by an extension of this interpretation to the nonresonant transport regime in Sec.\ \ref{subsubsec:1m_nonres}.

\begin{figure}[h!]
		\includegraphics[width=.6\textwidth]{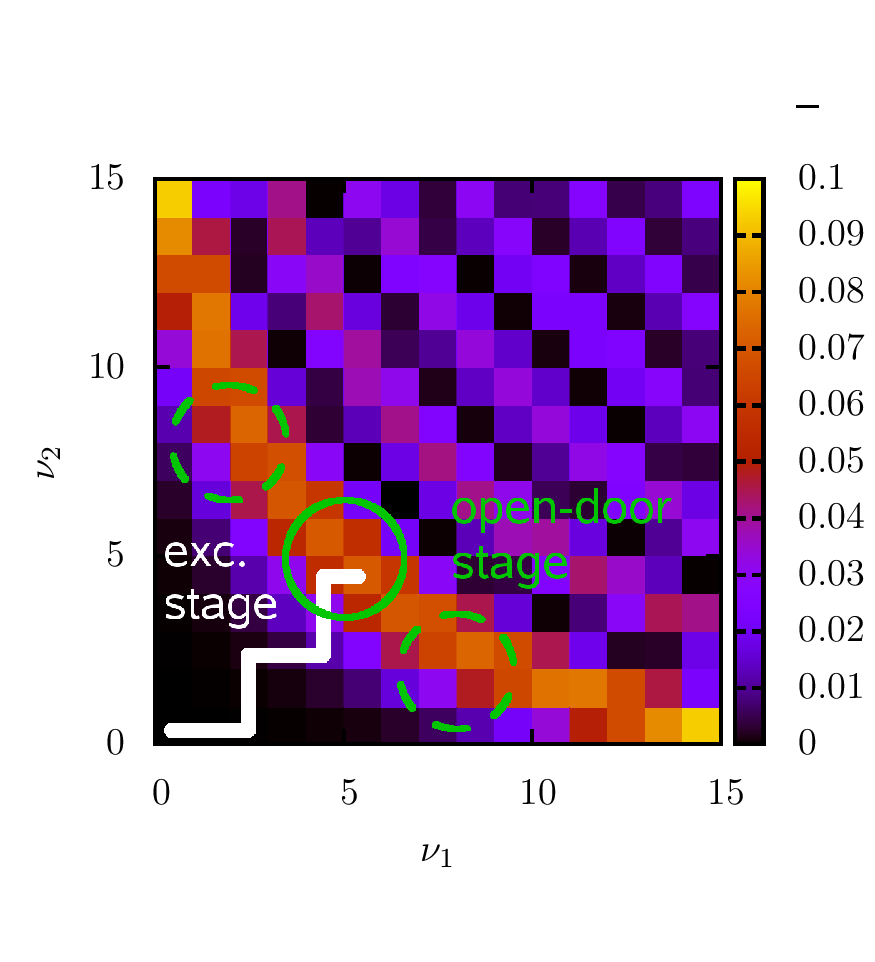}
\vspace{-1cm}
		\caption{FC-transition probabilities $|X_{v_1 v_2}|^2$ for $\lambda/ \Omega =4$. The excitation stage (white line) and the open-door stage (solid green circle) are shown for $\bar \epsilon_0=\Omega$ and $\Phi \in (2 \Omega, 4 \Omega)$. A configuration where two vibrational quanta can be excited during a transition across the right junction, whereof one facilitates the subsequent transition across the left junction. Additionally, the dashed green circle depicts the open-door stage for $\bar \epsilon_0=6 \Omega$ in the same bias voltage range.}
\label{fig:FC_1m_4lo_012e}
\end{figure}

To facilitate the subsequent discussion, we first consider the current-voltage characteristics for
the extended model depicted by the blue solid line in Fig.\ \ref{fig:4lo_I_F_el}a.  As is well known from previous studies,\cite{Koch2005,Koch2006} the strong coupling to the vibrational degrees of freedom results in a pronounced suppression of the current for low bias voltages and a stepwise increase of the current with bias voltage until it converges to the electronic value for $\Phi \gtrsim 2.5 \unit{V}$. The steps for voltages $\Phi \geq 2 (\bar \epsilon_0+ \Omega)$ are due to the successive contribution of vibronic transport processes. The strong suppression of the current, also
called Franck-Condon blockade (FC-blockade),\cite{Koch2005} can be attributed to the fact, that the Franck-Condon (FC-) transition probabilities $|X_{v_1 v_2}|^2$ between low-lying vibrational states ($v_1+v_2\lesssim 7$) of the unoccupied and the occupied molecular bridge, which are shown as colormap in Fig.\ \ref{fig:FC_1m_4lo_012e}, are exponentially suppressed.

%

\subsubsection{Avalanche-Like Transport in the Resonant Regime} \label{subsubsec:1m_res} 
The Fano factor-voltage characteristics for the model considered above is depicted by 
the solid blue line in Fig.\ \ref{fig:4lo_I_F_el}b.
We first concentrate on the resonant transport regime ($\Phi > 2 \bar \epsilon_0=2 \Omega$), especially on the bias voltage range $\Phi \in (2 \Omega,4 \Omega)$ where the Fano factor-voltage characteristics shows a local maximum of the order of $10^3$.
This super-Poissonian-value of the Fano factor ( $F>1$ ) is due to avalanche-like transport. Following Koch \emph{et. al.} \cite{Koch2005}, the term avalanche refers to periods of time where many electrons tunnel through the molecular bridge on a very short time scale. These periods are interrupted by significantly longer waiting times where no tunneling event occurs.
The typical course of such an avalanche is illustrated in Fig.\ \ref{fig:proc_1m_04lo}:
\begin{figure}[h!]
	\centering
\begin{tabular}{llll}
a)\hspace{1cm} & & b)\hspace{1cm} &\\
& 		\includegraphics[width=.17\textwidth]{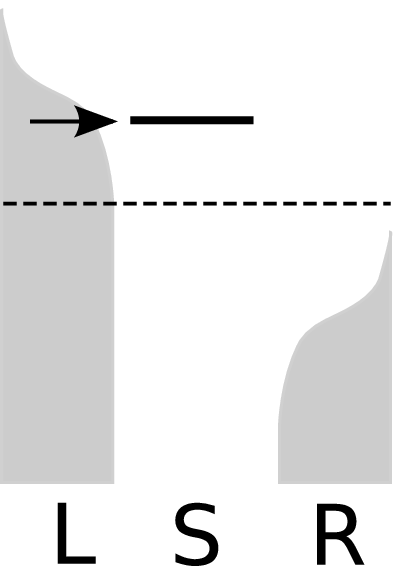}
& &
		\includegraphics[width=.17\textwidth]{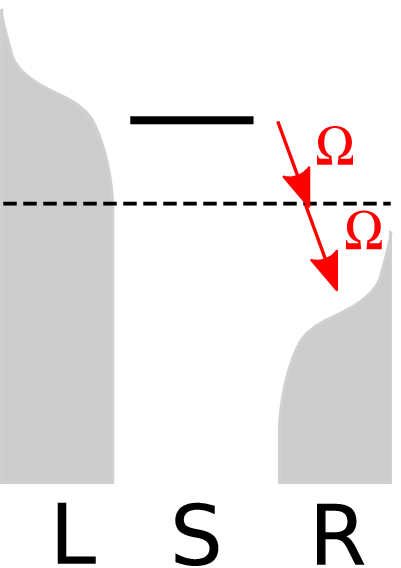}
\end{tabular}
		\caption{Illustration of the purely electronic trigger process (a) and the subsequent vibronic excitation process (b) in the resonant transport regime. Thereby, the shorthands $L$ and $R$ denote the left and the right lead, respectively and $S$ stands for the molecular bridge (system). The arrows illustrate tunneling processes, which are either purely electronic (black) or involve the excitation (red) of the molecular bridge by a vibrational quantum of the frequency $\Omega$.}
\label{fig:proc_1m_04lo}
\end{figure}
Assuming that the molecular bridge is unoccupied and in its vibrational ground state, an avalanche is triggered by a purely electronic transition from the left lead onto the molecule. This trigger process is depicted by a black arrow in Fig.\ \ref{fig:proc_1m_04lo}a. The probability for this ground to ground state transition $0 \stackrel{\tL \tS}{\longrightarrow}0$ is strongly suppressed by the Franck-Condon blockade. Thereby, the notation $v_1 \xrightarrow{\tL \tS / \tS \tR } v_2$ denotes a transition from the vibrational state $v_1$ to the state $v_2$ in a tunneling event from the left lead $L$ to the molecular bridge $S$ or from the molecular bridge to the right lead $R$, respectively (cf.\ Fig.\ \ref{fig:proc_1m_04lo}).
Due to the FC-transition probabilities $|X_{v_1 v_2}|^2$ (shown in Fig.\ \ref{fig:FC_1m_4lo_012e}) it is favorable to excite the molecular bridge by the maximum number of vibrational quanta (2 in this case) in the subsequent tunneling event from the molecule to the right lead.
The two red arrows in Fig.\ \ref{fig:proc_1m_04lo}b symbolize this process. For the subsequent transition across the left junction it is most probable again that an electron undergoes a purely electronic transition $2 \stackrel{\tL \tS}{\longrightarrow}2$ so that the molecular bridge can be excited by two additional vibrational quanta in a transition to the right lead. This corresponds to a net excitation of two quanta per tunneling cycle from the left to the right lead. We call this part of the avalanche ``excitation stage'', where the vibrational excitation of the molecular bridge is increased until purely electronic transitions $v \longrightarrow v$ are most probable ($v=4-5$ for $\lambda/\Omega=4$). At this excitation level many electrons  can tunnel through the molecular junction behind each other on a very short time scale due to the high FC-transition probabilities (cf. Fig.\ \ref{fig:FC_1m_4lo_012e}). We call this second part of the avalanche ``open-door stage''.
These two stages are indicated in Fig.\ \ref{fig:FC_1m_4lo_012e}. Thereby, the white line depicts the sequence of the most probable vibrational transitions the system undergoes in the excitation stage in order to reach the open-door stage. The open-door stage is centered around the electronic transitions $4 \to 4$ and $5 \to 5$ (marked by a solid green circle) so that it is not favorable for the system to obtain a higher vibrational excitation. This explanation is confirmed by the dashed green line in Fig.\ \ref{fig:4lo_I_F_el}b, which was obtained by restricting the vibrational basis in the calculation to 7 basis states, and which perfectly coincides with the blue solid line obtained with a converged vibrational basis.
Such an avalanche continues until the molecular bridge returns to the vibrational ground state $v \stackrel{\tS \tR}{\longrightarrow}0$ in an electronic tunneling event from the molecule to the right lead. This transition, which is strongly suppressed by the FC-blockade and thus very rare, is followed by a long waiting time until the next trigger process starts the whole procedure again.

These avalanches have already been analyzed in detail by Koch \emph{et al.}.\cite{Koch2005,Koch2005FCS,Koch2006,KochPhd} Assuming that the avalanches start and terminate in the vibrational ground state of the molecular bridge, they showed that the avalanches have a self-similar hierarchical structure. Based on the assumption that the typical duration of such an avalanche is much shorter than the waiting times in between, the Fano factor in the long-time limit can be expressed by \cite{Koch2005,KochPhd}
\begin{equation}
 F=\av{N_0} + \frac{\av{t_0^2} - \av{t_0}^2}{\av{t_0}^2} \av{N_0}.
\label{eq:F_Koch}
\end{equation}
Thereby, two contributions determine the Fano Factor: The average number of electrons $\av{N_0}$ in such an avalanche as well as the normalized variance $\left(\langle t_0^2 \rangle - \langle t_0 \rangle^2 \right) /\langle t_0 \rangle ^2$  of waiting times $t_0$ between the avalanches.
In the resonant transport regime an avalanche can terminate in two possible, nonequivalent charge states for $\bar \epsilon_0>\Omega / 2$: The electronic level is either occupied or unoccupied. For $\bar \epsilon_0=\Omega$ and $\Phi \in (2 \Omega, 4 \Omega)$ the situation can be described as follows: Assuming the molecular bridge is unoccupied, a new avalanche can start with the probability $|X_{00}|^2$ which is much smaller than the respective probability $|X_{02}|^2$ for the occupied case ($|X_{00}|^2 \ll |X_{02}|^2$). Consequently, the waiting time until the next avalanche depends on the charge state and thus $\left(\langle t_0^2 \rangle - \langle t_0 \rangle^2 \right) /\langle t_0 \rangle ^2>1$ holds, as shown by Koch \emph{et al.}.\cite{Koch2006}

For larger voltages, $\Phi > 4 \Omega$ in Fig.\ \ref{fig:4lo_I_F_el}b,
the Fano factor-voltage characteristics gradually decreases.
There are mainly two reasons for this behavior. On the one hand, the maximum net excitation per tunneling cycle, which is in general given by $[\text{mod}(\Phi/2-\bar \epsilon_0,\Omega) + \text{mod}(\Phi/2+\bar \epsilon_0,\Omega)]$ vibrational quanta in the resonant regime, increases with bias voltage. Consequently, the residence time of the system in the open-door stage is reduced, because the system can easily reach highly excited vibrational states ($v \gtrsim 13$), where a direct transition back in the vibrational ground state and thus the termination of the avalanche is favorable. We refer to these states as ``terminator'' states in the following. Consequently, the average number of electrons $\av{N_0}$ in an avalanche decreases and thus the Fano factor according to Eq.\ (\ref{eq:F_Koch}).
On the other hand, in the resonant transport regime the probability of the trigger process also increases with the bias voltage, so that the waiting times between the avalanches are shortened. According to our previous explanations, the absolute duration of the waiting times does not influence the Fano factor. However, this statement only holds as long as the waiting times $t_0$ between the avalanches are much longer than the duration of the avalanches itself.

In the high bias limit ($\Phi >24 \Omega$), the Fano factor assumes sub-Poissonian values, which correspond to anti-correlated transport due to the Pauli blockade as in the purely electronic case. This indicates that the system is no longer subject to FC-blockade as the blockade regime can be left within a single transition from the vibrational ground state.

\subsubsection{Avalanche-Like Transport in the Nonresonant Regime} \label{subsubsec:1m_nonres}
In the following, we analyze the Fano factor-voltage characteristics in the nonresonant transport regime. In this regime, which is the main focus of the remainder of this work, the Fano factor is exclusively determined by the average number $\av{N_0}$ of electrons within an avalanche, i.e.\ Eq.\ (\ref{eq:F_Koch}) reduces to
\begin{equation}
  F=2 \av{N_0}.
 \label{eq:F_Koch_2}
\end{equation}
This is due to the fact that both chemical potentials are below the energy of the electronic level in the nonresonant transport regime, so that the probabilities for a trigger process starting either from the unoccupied or the occupied molecular bridge, differ by orders of magnitude: The former process across the left junction, $0 \stackrel{\tL \tS}{\longrightarrow}0$, is suppressed by both the FC-blockade and the Fermi distribution in the left lead, whereas the latter process across the right junction, $0 \stackrel{\tS \tR}{\rightarrow} \text{mod}(\bar \epsilon_0+\Phi/2,\Omega)$, is only subject to moderate FC-blockade.
However, the zero-frequency noise can only probe the long-time limit, so that information on short timescales and thus on the inner fine structure of the avalanches is lost.\cite{Ubbelohde2012,Flindt2012}
Therefore, in terms of the zero-frequency noise it is sensible to assume that an avalanche is only terminated in the vibrational ground state of the unoccupied molecular bridge. As a result, the normalized variance of waiting times in Eq.\ (\ref{eq:F_Koch}) is equal to unity.

First, we consider the regime of small bias voltages ($\Phi < 2 \Omega=2 \bar \epsilon_0$), where the Fano factor-voltage characteristics shows a value of almost 20 (neglecting the singularity at $\Phi=0$). This super-Poissonian value of the Fano factor ($F>1$) is significantly smaller than the one obtained in the resonant transport regime above, which clearly indicates that the avalanches in this regime comprise notably fewer electrons on average. This is reflected in the typical course of such an avalanche in the regime $\Phi < 2 \Omega$, which is illustrated in Fig.\ \ref{fig:proc_1m_simav}:
\begin{figure}[h!]
	\centering
\begin{tabular}{llllll}
a)\hspace{1cm} & & b)\hspace{1cm} & & c)\hspace{1cm} &\\
& 		\includegraphics[width=.17\textwidth]{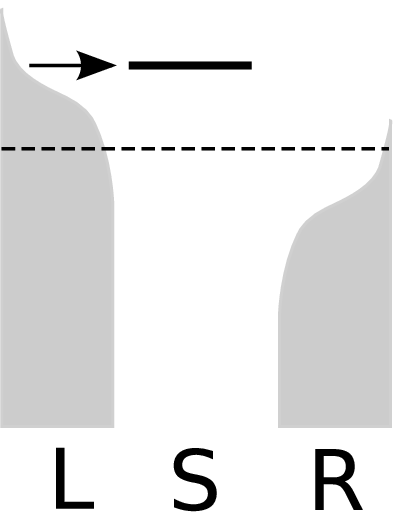}
& &
		\includegraphics[width=.17\textwidth]{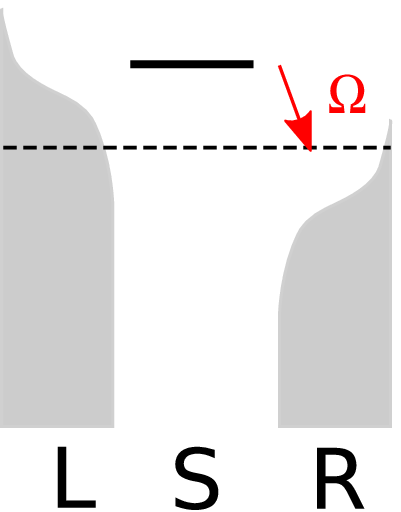}
& &
		\includegraphics[width=.17\textwidth]{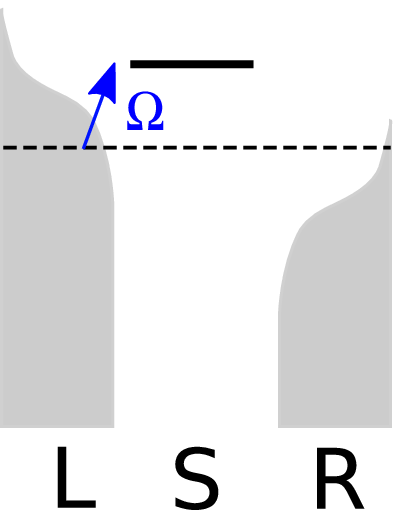}
\end{tabular}
		\caption{Illustration of the processes, which occur in the course of a ``simple'' avalanche. The arrows illustrate tunneling processes, which are either purely electronic (black) or comprise the excitation (red) / deexcitation (blue) of the molecular bridge by a vibrational quantum of the frequency $\Omega$.}
\label{fig:proc_1m_simav}
\end{figure}

 As both chemical potentials are below the energy of the electronic state, the trigger process $0 \stackrel{\tL \tS}{\longrightarrow}0$ is not only suppressed by the FC-blockade but it also requires the thermal broadening of the Fermi distribution in the left lead.\footnote{In a full description, including higher order corrections in the molecule-lead coupling, inelastic cotunneling processes may also act as additional trigger processes.} This process is sketched by a black arrow in Fig.\ \ref{fig:proc_1m_simav}a. However, the vibrational quantum which is excited in the subsequent transition $0 \stackrel{\tS \tR}{\longrightarrow}1$ across the right junction (cf.\ Fig. \ref{fig:proc_1m_simav}b) has to be reabsorbed by the molecular bridge in order to facilitate the subsequent electronic tunneling event from the left lead to the molecule $1 \stackrel{\tL \tS}{\longrightarrow}0$ (cf.\ Fig.\ \ref{fig:proc_1m_simav}c). This cycle of excitation and absorption continues until the molecular bridge returns to its vibrational ground state in a transition across the right junction, which is suppressed very strongly by the FC-blockade:
\begin{equation}
0 \stackrel{\tL \tS}{\longrightarrow} 0 \stackrel{\tS \tR }{\longrightarrow} 1 \stackrel{\tL \tS}{\longrightarrow}0 \stackrel{\tS \tR}{\longrightarrow} 1 \cdots 1 \stackrel{\tL \tS}{\longrightarrow} 0 \stackrel{\tS \tR}{\longrightarrow} 0.
\label{eq:sim_ava}
\end{equation}
We call these avalanches ``simple'' as there is no net excitation of the molecular bridge in a sequential tunneling cycle, i.e. the molecule is always in the vibrational ground state when one electron has tunneled from the left lead onto the molecule. As a result, for $\bar \epsilon_0=0.1 \unit{eV}$ the avalanches only involve the vibrational ground $\ket{0}$ and first excited state $\ket{1}$. This is confirmed by the black dashed line in Fig.\ \ref{fig:4lo_I_F_el}b, which corresponds to a calculation where the vibrational basis is restricted to the two lowest states $\ket{0}$ and $\ket{1}$, and which in the voltage range $\Phi < 2 \Omega$ coincides with the solid blue line obtained for a converged number of 100 vibrational basis states.

\begin{figure}[h!]
		\includegraphics[width=.8\textwidth]{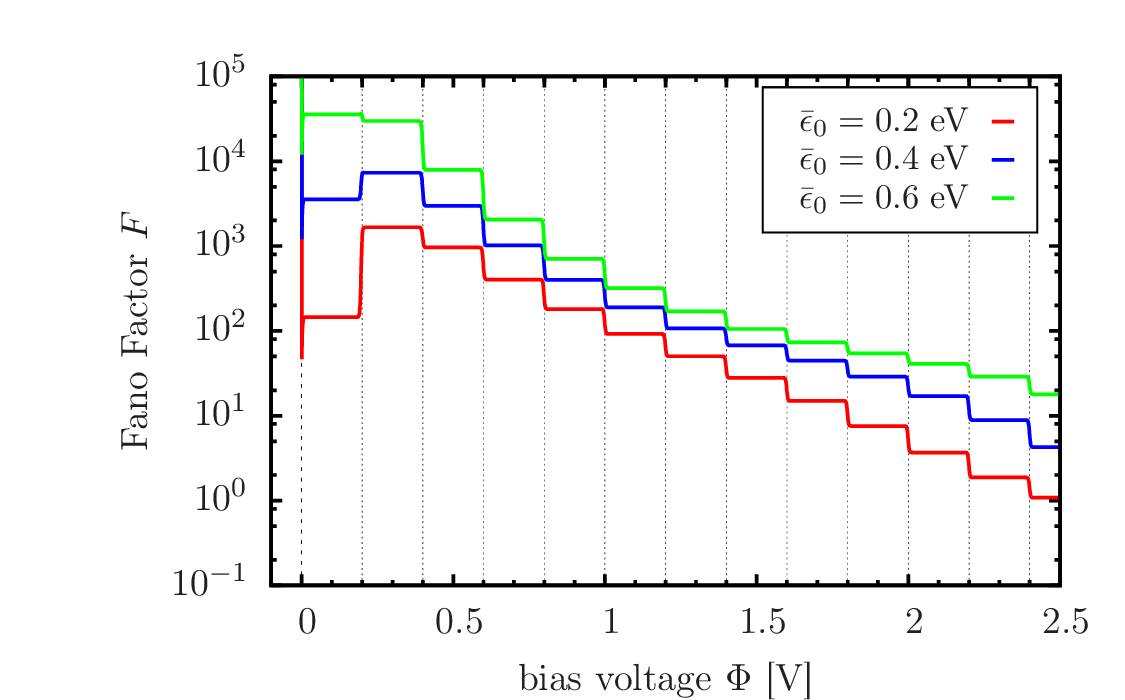}
		\caption{Fano factor-voltage characteristics in the strong coupling regime $\frac{\lambda}{\Omega}=4$ for different energies $\bar \epsilon_0$ of the electronic level at $T=10 \unit{K}$.
}
\label{fig:4lo_F_dife}
\end{figure}

In order to generalize the concept of ``simple'' avalanches, we consider in Fig.\ \ref{fig:4lo_F_dife} three additional systems with higher energies of the electronic level.
The results show that the Fano factor  increases with energy $\bar \epsilon_0$ in the complete voltage range shown here. Focussing again on the regime $\Phi <2 \Omega$, this behaviour can be explained by the generalization of the concept of ``simple'' avalanches to arbitrary values of $\bar \epsilon_0$:
\begin{equation*}
\setlength{\jot}{5pt}
0 \stackrel{\tL \tS}{\longrightarrow} 0 \stackrel{\tS \tR}{\longrightarrow} \begin{array}{c} 1 \\ \tiny{\vdots} \\ \nu \end{array} \stackrel{\tL \tS}{\longrightarrow}0 \stackrel{\tS \tR}{\longrightarrow} \begin{array}{c} 1 \\ \tiny{\vdots} \\ \nu \end{array} \cdots \begin{array}{c} 1 \\ \tiny{\vdots} \\ \nu \end{array} \stackrel{\tL \tS}{\longrightarrow}0 \stackrel{\tS \tR}{\longrightarrow} 0.
\end{equation*}
The molecular bridge can be excited by a maximum of $\nu=\text{mod}(\bar \epsilon_0,\Omega)$ vibrational quanta within a tunneling event across the right junction. If only $(1, \cdots, \nu-1)$ vibrational quanta are excited in such a transition, the subsequent transition across the left junction has to be facilitated by the thermal broadening of the Fermi distribution.
However, as stated above, the zero-frequency noise cannot resolve the inner fine structure of an avalanche.
Consequently, the longest time scale is given by the waiting time between the avalanches, which are triggered by the transition $0 \stackrel{\tL \tS}{\longrightarrow}0$ whereas all other transitions starting from higher excited states $(1,\dots, \nu) \stackrel{\tL \tS}{\longrightarrow}0$ appear simultanously in the zero-frequency picture.
As a result, the ratio $\sum_{i=1}^\nu |X_{0i}|^2/|X_{00}|^2$ determines the average number of electrons in such an avalanche, which is terminated by the transition $0 \stackrel{\tS \tR}{\longrightarrow}0$ across the right junction. This ratio rises with $\bar \epsilon_0$ and thus the Fano factor also increases.

Next, we concentrate on avalanches which can reach the 'open-door' stage in the nonresonant regime. To this end, we analyze the voltage interval $\Phi \in (2 \Omega,4 \Omega)$ where the Fano factor-voltage characteristics shows a local maximum for $\bar \epsilon_0 \leq 0.4 \unit{eV}$ as can be seen from Figs.\ (\ref{fig:4lo_I_F_el},\ref{fig:4lo_F_dife}).
This maximum can be attributed to the fact, that in this voltage range only a maximum net excitation of 2 vibrational quanta can be achieved in a tunneling cycle from the left to right lead.
As already discussed for the resonant regime above, this small net excitation prevents the system from reaching the highly excited ``terminator'' states ($v\gtrsim 13$) and thus prolongs the avalanches. 

To understand, why the Fano factor increases with the energy $\bar \epsilon_0$ of the electronic level in the nonresonant regime, it is important to notice, that in contrast to the resonant regime the electrons cannot tunnel in purely electronic tunneling cycles through the system because both chemical potentials are below the energy of the electronic level.
Therefore, tunneling cycles have to incorporate the excitation and absorption of vibrational quanta:
%
At least $\lceil ( \bar \epsilon_0 - \Phi /2 ) / \Omega \rceil=\text{mod}(\bar \epsilon_0,\Omega)-1$ vibrational quanta have to be absorbed in order to facilitate a transition across the left junction.
As a result, with increasing energy of the electronic level $\bar \epsilon_0$ the open-door stage is shifted further away from the regime of purely electronic transitions $v \to v$ along the branches of the FC-parabola to transitions $v \leftrightarrow v - (\text{mod}(\bar \epsilon_0,\Omega)-1)$.
For $\bar \epsilon_0=6 \Omega$, this is illustrated by the two dashed green circles in Fig.\ \ref{fig:FC_1m_4lo_012e} which are centered around the most probable transitions. 
That means that after a tunneling event across the left junction the average vibrational excitation of the molecular bridge is reduced with increasing $\bar \epsilon_0$.
However, direct transitions from these low-lying vibrational states  into the vibrational ground state and thus the termination of the avalanche are suppressed due to FC-blockade for $\frac{\lambda}{\Omega}=4$.
Consequently, the average number of electrons in such an avalanche, which runs in the open-door stage, increases with $\bar \epsilon_0$ and thus also the Fano factor. 
%

\subsection{Multimode Vibrational Effects} \label{sec:results_mm}
\begin{figure}[htbp]
  \subfigure[$\Delta Q=1$]{
    \label{Labelname 1}
    \includegraphics[width=.8\textwidth]{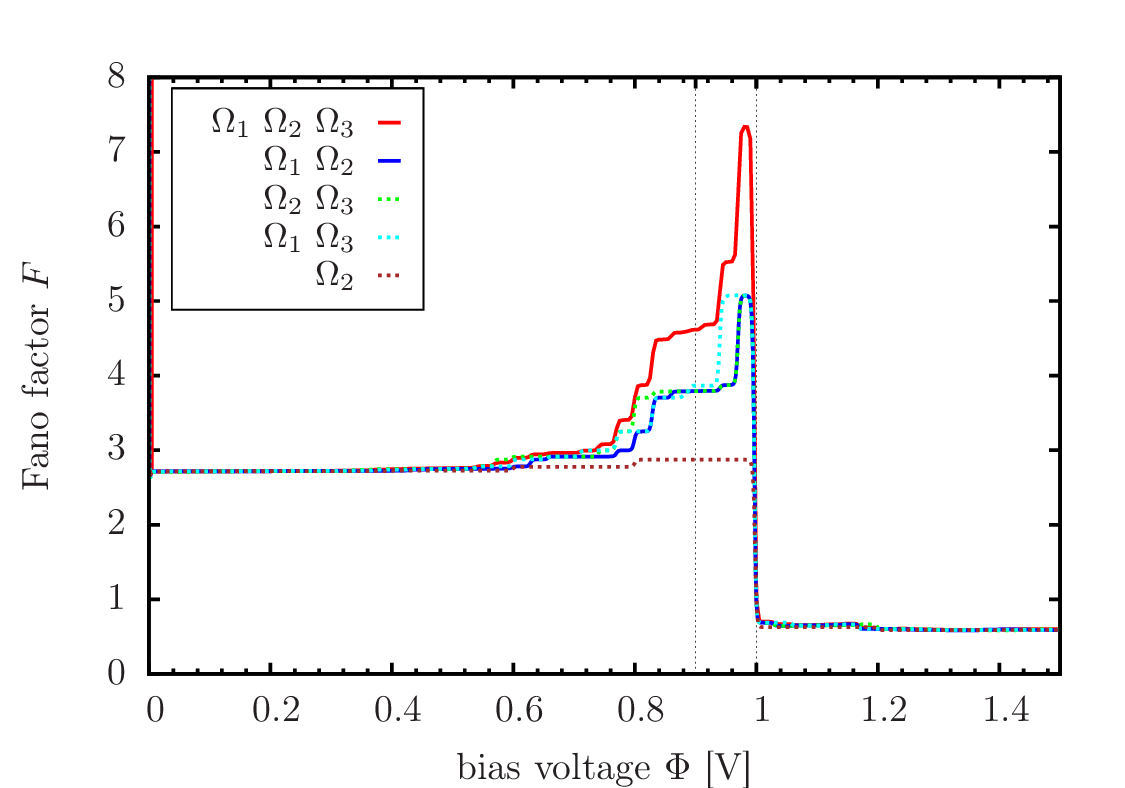}
\vspace{-0.5cm}
  }
  \subfigure[$\Delta Q=3$]{
    \label{Labelname 2}
    \includegraphics[width=.87\textwidth]{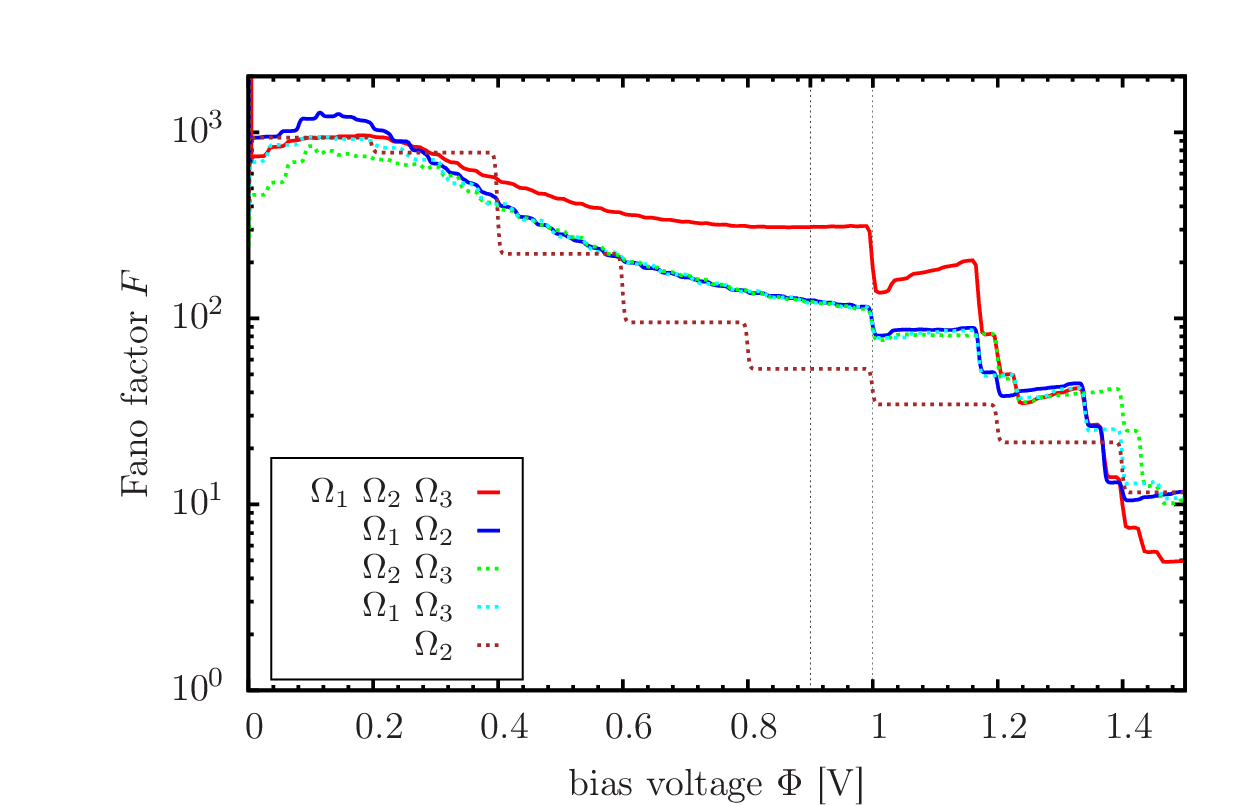} \hspace{0.9cm}
\vspace{-0.5cm}
  }
  \caption{Fano factor $F$ as a function of bias voltage $\Phi$ for the electronic-vibrational coupling to 1-3 vibrational modes with frequencies $\Omega_1=85 \unit{meV}$, $\Omega_2=100 \unit{meV}$ and $\Omega_3=115 \unit{meV}$ calculated at $T=10\unit{K}$ and $\Gamma=0.2\unit{meV}$. The respective energy of the electronic level is given by $\bar \epsilon_0=0.5 \unit{eV}$. The panels (a) and (b) correspond to different values of the dimensionless shift $\Delta Q$.}
  \label{fig:3m_F_V_difQ}
\end{figure}
All systems considered so far involve only a single vibrational mode. Realistic molecular junctions, however, typically have a multitude of modes and thus represent significantly more complex systems. In this section, we extend our analysis and investigate vibrational effects in the 
noise due to electronic-vibrational coupling to several vibrational modes.
Thereby, three different model systems, which differ by the distance of the single electronic state from the Fermi level, are studied at the onset of the resonant current and multimode effects in the shot noise signal are discussed. The corresponding parameters are summarized in Tab.\ \ref{tab:mm_models}.


%
 \begin{table}
 \centering
  \begin{tabular}{c | c c c c c c c}
  Model & $\bar \epsilon_0$ [eV] & $\Phi$ [V] & $\Omega_1$ [meV] &  $\Omega_2$ [meV] & $\Omega_3$ [meV] & $\Gamma$ [meV] & $T$ [K] \\ \hline
  1 & 0.08 & 0.12 & 85 & 100 & 115 & 0.2 & 10\\
  2 & 0.14 & 0.2 & 85 & 100 & 115 & 0.2 & 10\\
  3 & 0.5 & 0.9 & 85 & 100 & 115 & 0.2 & 10
 \end{tabular}
 \caption{Summary of the parameters for model 1-3.}
 \label{tab:mm_models}
 \end{table}
%
%
%
\subsubsection{Dependence of the Fano Factor on Bias Voltage} \label{sec:FV} 

Fig.\ \ref{fig:3m_F_V_difQ} shows the Fano factor-voltage characteristics for models with up to 
three vibrational modes.
To allow a meaningful comparison of results for a different number of vibrational modes, the dimensionless displacement $\Delta Q=\sqrt{\sum_{\alpha}\lambda_{\alpha}^{2}/\Omega_{\alpha}^{2}}$ 
between the minima of the potential energy surfaces of the occupied and the unoccupied molecule is chosen to be the same for all systems considered. Furthermore, the respective dimensionless couplings $\lambda_{\alpha}/\Omega_{\alpha}$ of all involved modes are chosen to be identical so that the coupling constants $\lambda_\alpha=\Omega_\alpha \Delta Q / \sqrt{M} $ are determined by the number $M$ of involved vibrational degrees of freedom with frequencies $\Omega_\alpha$ and the respective dimensionless shift $\Delta Q$.

We first consider the Fano factor-voltage characteristics in Fig.\ \ref{fig:3m_F_V_difQ}a obtained for $\bar \epsilon_0=0.5 \unit{eV}$ and a moderate dimensionless shift $\Delta Q=1$. For small bias voltages ($\Phi <0.5 \unit{V}$) the Fano factor-voltage characteristics for a different number of vibrational modes almost coincide at a small super-Poissonian value and exhibit a steplike increase for higher bias voltages until they reach their maximum value ($<10$) directly before the onset of resonant transport ($\Phi\approx 2 \bar \epsilon_0$).
Thereby, the number of steps increases with the number of involved vibrational modes because each step corresponds to the emergence of a new vibronic transport channel.
In models with multiple vibrational modes, a significant part of these transport channels is associated with multimode processes, where the vibrational excitation of
several vibrational modes is changed within one tunneling event.
In the resonant transport regime ($\Phi > 2 \bar \epsilon_0$) the Fano factors are sub-Poissonian ($F<1$).
%

Next, we analyze the Fano factor-voltage characteristics for a  dimensionless shift $\Delta Q=3$ (cf.\ Fig.\ \ref{fig:3m_F_V_difQ}b), corresponding to a larger electron-vibrational coupling in the regime of FC-blockade. Similar as for a single vibrational more (cf.\ Sec.\ \ref{subsubsec:1m_nonres}), the Fano Factor-voltage characteristics exhibits a maximum, which is of the order of $10^3$ for $\Phi \in (0,0.2 \unit{V})$ in the nonresonant regime. While the curves obtained for a different number of vibrational modes are of the same order of magnitude for small bias voltages ($\Phi \in (0,0.4 \unit{V})$), they differ significantly before the onset of resonant transport. The brown line corresponding to a single vibrational mode is almost one order of magnitude below the three-mode curve and the blue and the green line refering to the coupling to two modes are in between. In the following we will analyze the behavior in the voltage range directly before the onset of resonant transport in more detail. This voltage range is also particularly interesting because of the experimental findings of Secker \emph{et al.}, who reported very large shot noise levels directly before the onset of the resonant transport regime.\cite{Secker2011}

%
\subsubsection{Dependence of the Fano Factor on Coupling} \label{sec:FQ}

\begin{figure}[htbp]
    \label{Labelname 1}
    \includegraphics[width=.8\textwidth]{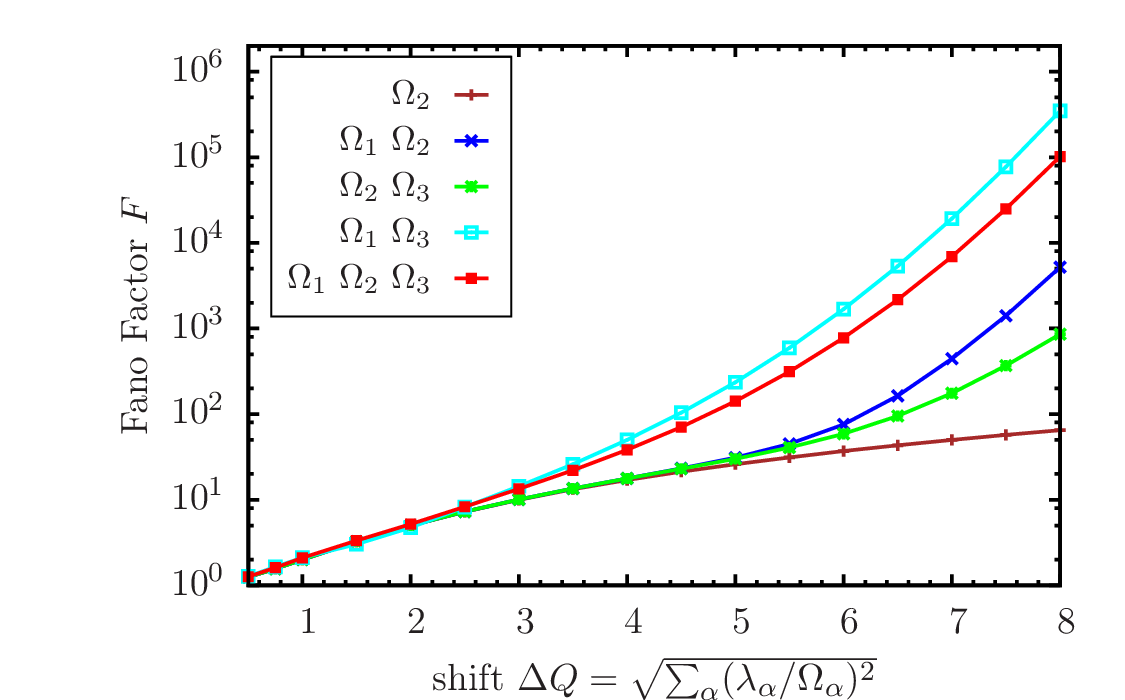}
  \caption{Fano factor $F$ as a function of the dimensionless shift $\Delta Q$ for model 1 at $T=10\unit{K}$. Results obtained for a different number of vibrational modes are shown and they are labeled by the frequencies $\Omega_\alpha$ of the involved modes.}
  \label{fig:3m_F_012V_008e}
\end{figure}

In order to analyze the Fano factor-voltage characteristics at the onset of the resonant transport regime ($\Phi \in (1.6 \bar \epsilon_0,2 \bar \epsilon_0)$) in more detail, we consider the Fano factor as a function of the dimensionless shift $\Delta Q$.
To this end, we study three different positions of the electronic level corresponding to model 1-3 (cf.\ Tab.\ \ref{tab:mm_models}). 
The transport mechanisms in these three model systems differ by the number of vibrational excitation and absorption processes which can occur.
%
%
\paragraph{Electronic State Close to Fermi Level.}
We start with model 1, which illustrates the fundamental relevance of multimode vibrational processes for the shot noise in general.
The energy of the electronic level in this model is chosen close to the Fermi level as $\bar \epsilon_0=0.08\unit{eV}$ such that it fulfills the relation $\Omega_{\alpha}/2< \bar \epsilon_{0}<\Omega_{\alpha}$. The corresponding results are shown in Fig.\ \ref{fig:3m_F_012V_008e}. Thereby, a voltage of $\Phi=0.12\unit{V}$ is considered to ensure that only a minimum number of processes can occur. Specifically, for all involved modes the molecular bridge can only be excited by a single vibrational quantum in a transition across the right junction and only one vibrational quantum has to be absorbed by the tunneling electron in order to facilitate a resonant transition from the left lead onto the molecule. 
Without multimode processes the avalanches cannot reach the open-door stage. Consequently, only ``simple'' avalanches, involving  vibrational ground and first excited state, can occur for the system with a single vibrational mode represented by the brown line. The results for two and three modes follow this trend for small to moderate coupling, $\Delta Q\lesssim 2.5$, which shows that ``simple'' avalanches dominate in this regime.

\begin{figure}[h!]
	\centering
\begin{tabular}{llllllll}
a)\hspace{0.5cm} & & b) \hspace{0.5cm} & & c)\hspace{0.5cm} & & d)\hspace{0.5cm} &\\
& 		\includegraphics[width=.18\textwidth]{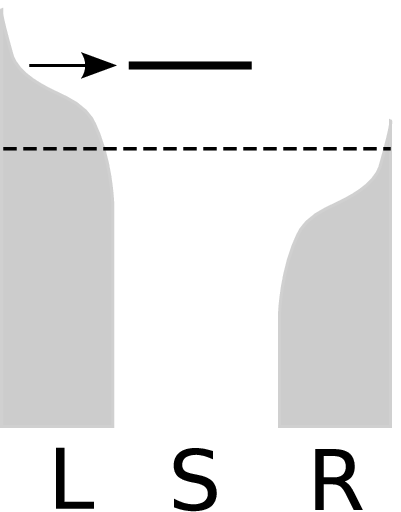}
& &
		\includegraphics[width=.18\textwidth]{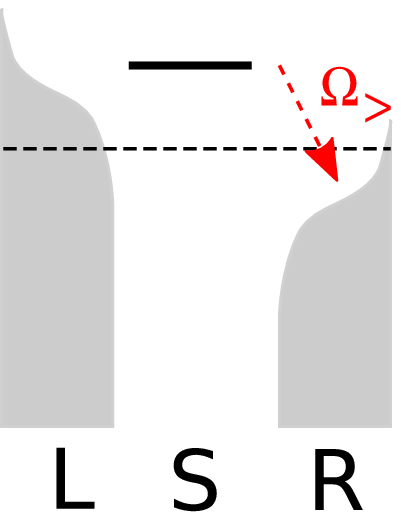}
& &
		\includegraphics[width=.18\textwidth]{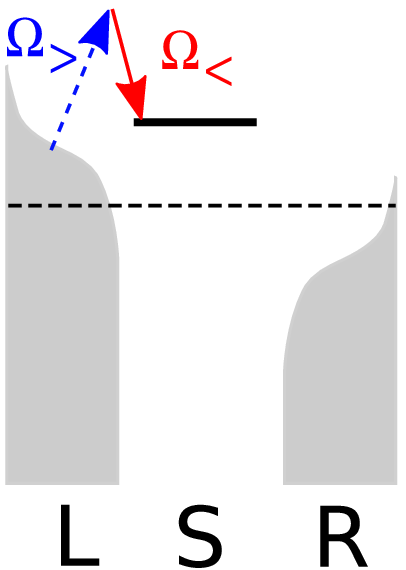}
& &
		\includegraphics[width=.18\textwidth]{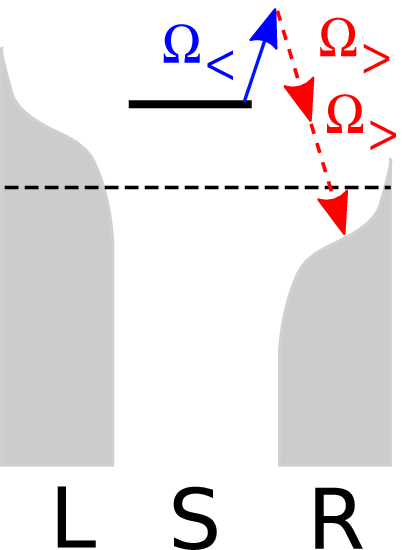}
\end{tabular}
		\caption{Illustration of the crucial multimode process in model 1. Red (blue) arrows correspond to excitation (absorption) of one vibrational quantum, where the dashed arrows refer to the vibrational mode with the higher frequency $\Omega_>$.}
\label{fig:proc_3m_012e}
\end{figure}

This behavior changes for larger couplings. Fig.\ \ref{fig:3m_F_012V_008e} shows that for dimensionless shifts $\Delta Q \gtrsim 2.5$ the Fano factor for three modes(red line) as well as in case of the two modes $\Omega_1$ and $\Omega_3$ (cyan line) increase much stronger with the coupling than for the other two- and one-mode systems.
The analysis shows that this is due to multimode processes, i.e.\ tunneling processes where the vibrational excitation of more than a single mode is changed. These processes become active in the regime of FC-blockade ($\Delta Q \gtrsim 2$), where with increasing $\Delta Q$ the vibrational ground to ground state transition $0 \rightarrow 0$ becomes more suppressed and processes comprising more and more vibrational quanta become favorable. Fig.\ \ref{fig:proc_3m_012e} illustrates the crucial multimode process for both cases:
After the trigger process (cf.\ Fig.\ \ref{fig:proc_3m_012e}a), the molecular bridge can be exited by one vibrational quantum of the high frequency mode ($\Omega_> \equiv \Omega_3=115 \unit{meV}$) (cf.\ Fig.\ \ref{fig:proc_3m_012e}b). In a next step, this quantum can be absorbed by a tunneling electron while one quantum of the low-frequency mode ($\Omega_< \equiv\Omega_1=85 \unit{meV}$) is generated simultaneously in a tunneling event across the left junction (cf.\ Fig.\ \ref{fig:proc_3m_012e}c). This is possible, because the difference between the left chemical potential and the electronic level is 20 meV and thus smaller than the difference between the highest and the lowest frequency.
The thus generated quantum of the low-frequency mode can now be absorbed in a subsequent tunneling event across the right junction and consequently two quanta of the medium frequency mode can be emitted (cf.\ Fig.\ \ref{fig:proc_3m_012e}d).
%
%
%
This multimode process facilitates a net vibrational excitation per tunneling cycle from the left to the right lead, so that avalanches can be initiated which reach the open-door stage, where many electrons can tunnel through the molecular junction on a very short time scale until the avalanche terminates. As already mentioned in Sec.\ \ref{subsubsec:1m_nonres}, this type of avalanche typically comprises a larger number of electrons than the ``simple'' avalanches occuring in the one mode case and thus leads to significantly higher Fano factors.

At first sight, it seems to be astonishing that the cyan curve in Fig.\ \ref{fig:3m_F_012V_008e}, corresponding to two modes, exceeds the three mode curve for shifts $\Delta Q >3 $. However, because only vibrational modes with frequencies $\Omega_1$ and $\Omega_3$ are involved in the crucial multimode processes leading to a net vibrational excitation (cf.\ Fig.\ \ref{fig:proc_3m_012e}), the presence of the vibrational mode $\Omega_2$ only reduces the probability for the net excitation.
Indeed, if the vibrational mode with frequency $\Omega_2$ is replaced by a mode with a slightly higher frequency $\Omega_2'=110 \unit{meV}$ in the three mode case, the Fano factor-voltage characteristics obtained for this frequency combination exceeds the cyan curve (data not shown). This can be attributed to fact that the frequency pairs  $\Omega_3$ and $\Omega_1$ as well as $\Omega_2'$ and $\Omega_1$ fulfill the relation $\Omega_3 - \Omega_1, \Omega_2' - \Omega_1> \bar \epsilon_0- \Phi/2$ so that the crucial multimode process described above can be facilitated  by both frequency pairs.

It is also interesting to note that those two-mode curves in Fig.\ \ref{fig:3m_F_012V_008e}, which represent the frequency pairs $\Omega_1$ and $\Omega_2$ as well as $\Omega_2$ and $\Omega_3$, do not deviate from the one-mode curve until the shift $\Delta Q$ has reached a value of 4. In principle, multimode processes cannot occur in this energy-voltage regime as the energy difference between the two involved modes is smaller by $5 \unit{meV}$ than the energy gap between the left chemical potential and the electronic level.
However, based on the fact that for this high value of $\Delta Q$ the vibrational ground to ground state transition is strongly suppressed, the same multimode process described above for the three-mode case is facilitated by the thermal broadening of the Fermi distribution. Thereby, an electron in the left lead with an energy of 65 meV absorbs one quantum of the high frequency mode and simultaneously emits one quantum of the low frequency mode. 
This hypothesis is confirmed by the fact that the onset of these multimode processes is shifted to larger $\Delta Q$ for smaller temperatures, i.e.\ it explicitly depends on the value of the Fermi distribution in the left lead at an energy of 65 meV (data not shown).

It is also seen that  the two-mode curve (blue line) representing the pair of lower frequencies increases more strongly than the green curve corresponding to the pair of higher frequencies. Due to the lower frequencies involved, absorption and emission processes can comprise more vibrational quanta. This favors the diversity of these avalanches and thus leads to a higher mean number of electrons within such an avalanche.

Another important observation is that the slope of the Fano factor-voltage characteristics for one vibrational degree of freedom (represented by the solid brown line in Fig.\ \ref{fig:3m_F_012V_008e}) gradually decreases  with increasing dimensionless shift $\Delta Q$. In contrast, the slope of the two- and three-mode curves increases with $\Delta Q$ after multimode effects have become active. This different behavior is due to the fact that avalanches only comprise the alternating emission and absorption of one vibrational quantum in the one-mode case. The only effect that occurs with increasing shift is that the ratio between the probability for these two processes, $|X_{01}|^2$, and the breakdown process $|X_{00}|^2$ increases quadratically in the shift, i.e. that $|X_{01}|^2/|X_{00}|^2=(\Delta Q)^2$. This influences the Fano factor much less than in the two and three mode case, where a net excitation per tunneling cycle is possible. This net excitation facilitates many different possibilities for transitions in the open-door stage, whose probabilities strongly depend on the dimensionless shift $\Delta Q$.
%
\paragraph{Intermediate Distance Between Electronic State and Fermi Level.}
We next consider model 2 with an electronic energy level of $\bar \epsilon_0=0.14\unit{eV}$ that the fulfills the relation $\Omega_\alpha <\bar \epsilon_0 < 2 \Omega_\alpha$. At bias voltage $\Phi=0.2 \unit{V}$ this model system represents a parameter regime where multimode effects do not influence the shot noise level significantly.
In contrast to model 1, two vibrational quanta can be excited in a tunneling event across the right junction where only one of them has to be absorbed by a tunneling electron to facilitate the subsequent transition across the left junction. This means that a net excitation of one vibrational quantum is possible in each tunneling cycle from the left to the right lead even if multimode effects are not taken into account.

\begin{figure}[h!]
		\includegraphics[width=.8\textwidth]{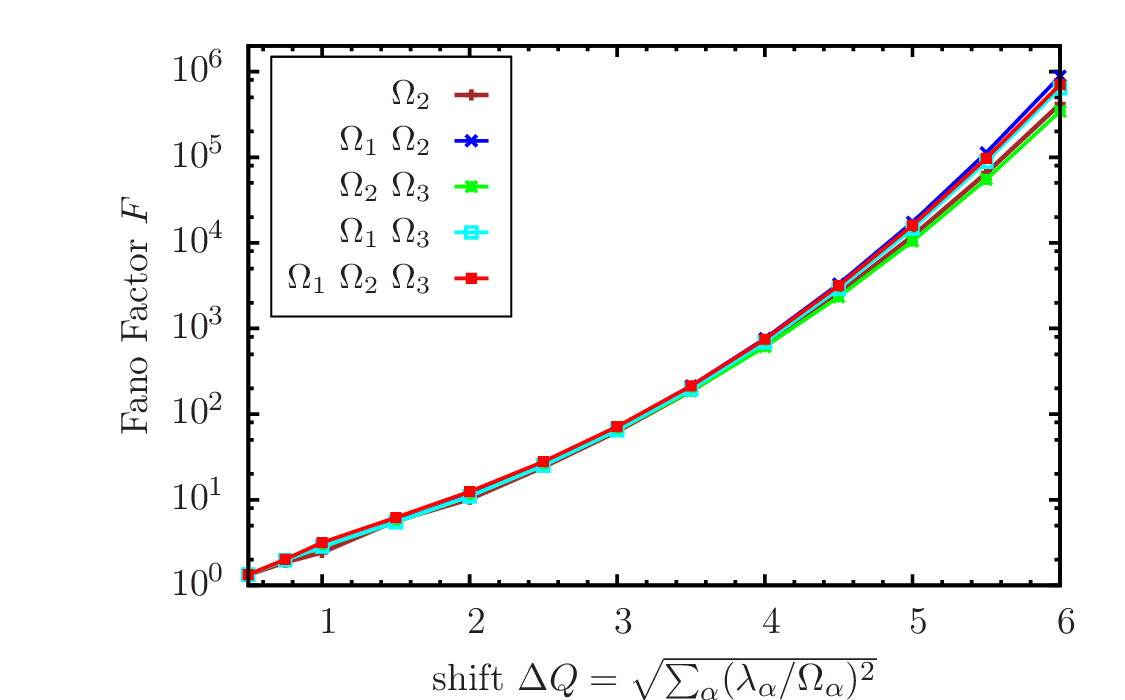}
		\caption{Fano factor $F$ as a function of the dimensionless shift $\Delta Q$ for model 2 at $T=10\unit{K}$.}
\label{fig:3m_F_02V_014e}
\end{figure}
%
%
For $\Delta Q \lesssim 4$ the five $F$-$\Delta Q$ characteristics in Fig.\ \ref{fig:3m_F_02V_014e} obtained for a different number of modes and frequency combinations almost coincide. This suggests that in this parameter regime multimode processes do not influence the shot noise level significantly.
This can be rationalized by the following two arguments: First, the avalanches can reach the open-door stage even without multimode processes in this regime. Secondly, due to the fact that the involved vibrational modes $\Omega_1, \Omega_2, \Omega_3$ are quite close in frequency ($| \Omega_\alpha -\Omega_\beta| \ll \Omega_\gamma$) and that the electronic level $\bar \epsilon_0$ is still relatively close to the Fermi energy $E_F$, only few multimode processes can occur which emerge from the frequency difference of the involved vibrational modes.
Consequently, the results in this regime do not depend very sensitively on the explicit choice of the frequencies of the involved vibrational modes.

Only for very large coupling strengths, $\Delta Q\gtrsim 4.5$, which are rarely found in molecules, the curves corresponding to a different number of vibrational modes and combination of frequencies start to deviate from each other indicating that multimode processes become important.

\paragraph{Electronic State Located Remote from Fermi Level.}
In order to interpret the behavior of the Fano factor in the nonresonant regime for the case that the electronic state is located remote from the Fermi level, we consider model 3 (cf.\ Tab.\ \ref{tab:mm_models}) and analyse its dependence on the dimensionless shift $\Delta Q$ at the bias voltage $\Phi=0.9 \unit{V}$ (marked by the left dashed horizontal line in Fig.\ \ref{fig:3m_F_V_difQ}).

\begin{figure}[h!]
		\includegraphics[width=.8\textwidth]{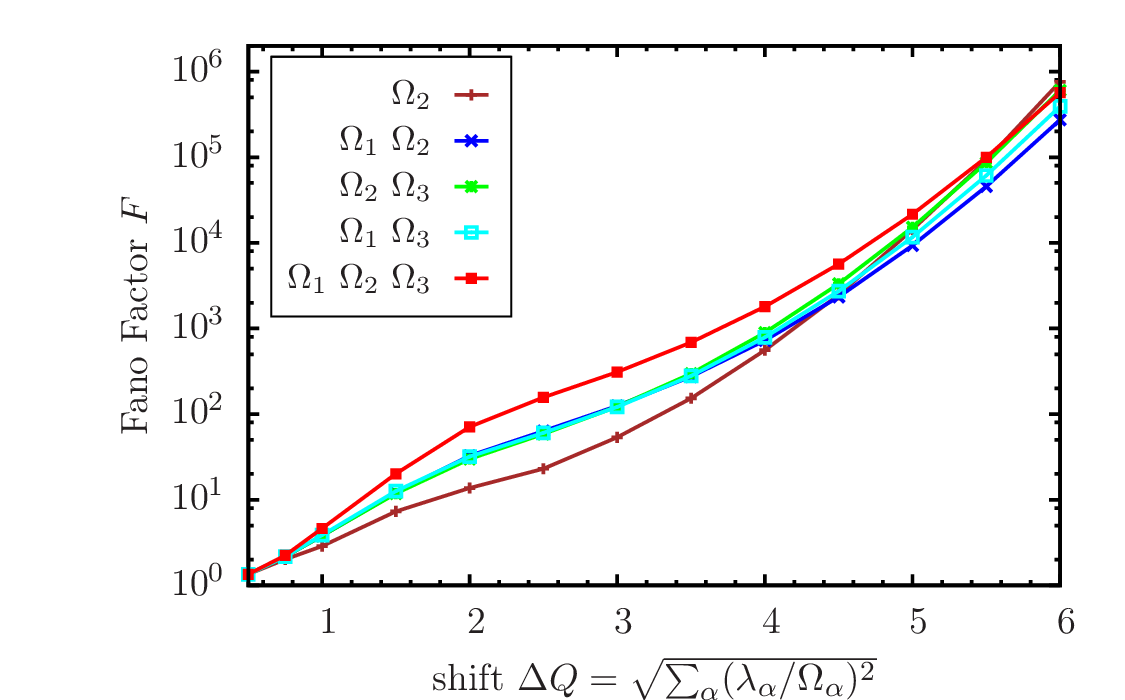}
		\caption{Fano factor $F$ as a function of the dimensionless shift $\Delta Q$ for model 3 at $T=10\unit{K}$.}
\label{fig:3m_F_09V_05e}
\end{figure}

Because the electronic state in model 3 is located remote from the Fermi level, many more excitation and deexcitation processes can occur than for model 1 or 2, respectively.
Fig.\ \ref{fig:3m_F_09V_05e} shows the corresponding curves for the electronic-vibrational coupling to 1-3 modes.
For $\Delta Q<4$ there is a clear hierarchy in the $F$-$\Delta Q$ characteristics: The red line corresponding to the three mode case shows the highest Fano factor whereas the almost identical blue, green and cyan curves obtained for two vibrational modes are located below. The one-mode case represented by the brown graph line exhibits an even smaller Fano factor.

\begin{figure}[h!]
		\includegraphics[width=.8\textwidth]{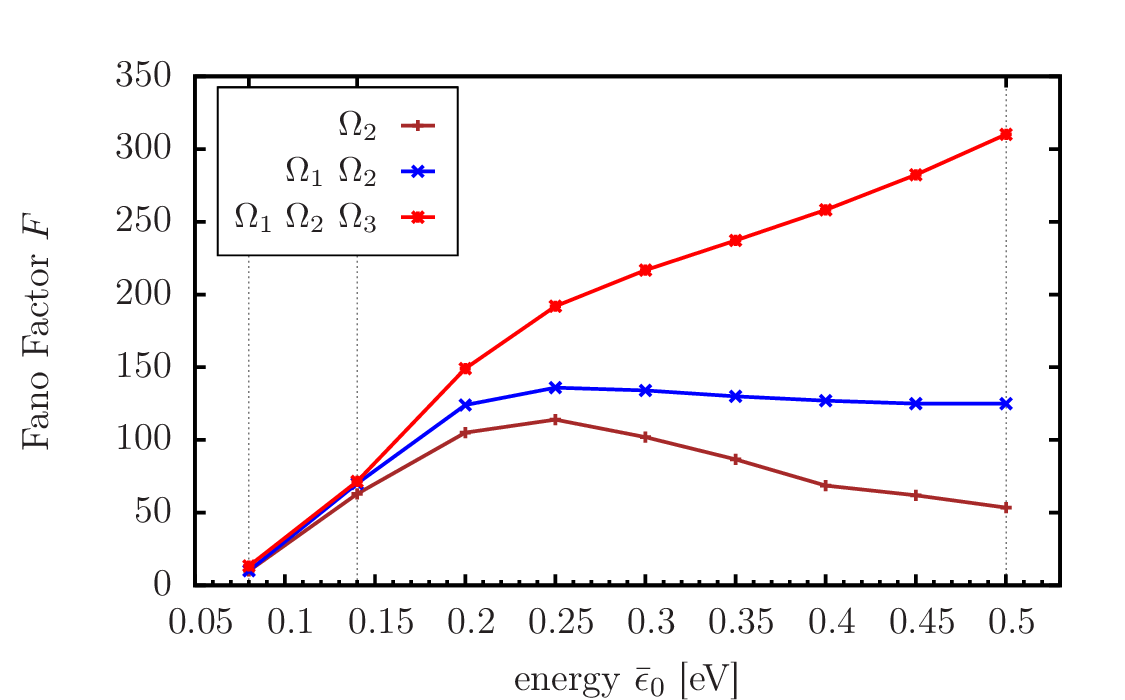}
\caption{Fano factor $F$ as a function of the energy $\bar \epsilon_0$ of the electronic level for the constant dimensionless shift $\Delta Q=3$. The results are shown for the coupling to one, two and three vibrational modes.
The data points marked by the vertical dashed lines correspond to the parameters of model 1 ($\bar \epsilon_0=0.08 \unit{eV}$, $\Phi=0.12 \unit{V}$), model 2 ($\bar \epsilon_0=0.14 \unit{eV}$, $\Phi=0.2 \unit{V}$) and model 3 ($\bar \epsilon_0=0.5 \unit{eV}$, $\Phi=0.9 \unit{V}$). The bias voltage for all other points is obtained according to the rule $\Phi=2 \bar \epsilon_0 -0.1$.}
\label{fig:3m_F_dife}
\end{figure}

In the following, we analyze the Fano factor voltage characteristics in the most interesting regime  $\Delta Q \in (2,3.5)$ in more detail. 
%
To this end, Fig.\ \ref{fig:3m_F_dife} shows the Fano factor  as function of the energy of the electronic level $\bar \epsilon_0$ for one, two and three vibrational modes at $\Delta Q=3$. Thereby, the corresponding bias voltage is given by $\Phi=2 \bar \epsilon_0 -0.1$ except for
 the first two points, where the voltage corresponds to the parameters of model 1 ($\bar \epsilon_0=0.08 \unit{eV}$ and $\Phi=0.12 \unit{V}$) and model 2 ($\bar \epsilon_0=0.14 \unit{eV}$ and $\Phi=0.2 \unit{V}$), respectively.
Focussing on the brown one-mode curve first, we observe an increase of the Fano factor with the energy $\bar \epsilon_0$ of the electronic level until it reaches its maximum around $\bar \epsilon_0=0.25 \unit{eV}$. This behavior is due to the fact that the net excitation of one vibrational quantum appearing in model 2 is not sufficient to reach the optimum open-door stage but a net excitation of three quanta is required. For $\bar \epsilon_0 >0.25 \unit{eV}$ the Fano factor decreases again: With increasing net excitation the system can probe many states and consequently the transitions are not restricted to the open-door stage anymore. As a result, the avalanche can reach highly excited states where a direct transition into the vibrational ground state and thus the termination of the avalanche is favorable ("terminator" states).

Next, we consider the blue two-mode curve representing the vibrational modes $\Omega_1=85 \unit{meV}$ and $\Omega_2=100 \unit{meV}$. This curve exceeds the one-mode curve for all values of $\bar \epsilon_0$ and remains almost constant for $\bar \epsilon_0 \geq 0.25 \unit{eV}$. This behavior can be attributed to multimode vibrational processes which lift the suppression of the Fano factor for large $\bar \epsilon_0$, where up to 11 vibrational quanta can be excited in a transition from the molecule to the right lead for $\epsilon_0=0.5 \unit{eV}$:
Due to these multimode processes both pairs of involved modes can reach even higher excited states than the more strongly coupled single mode $\Omega_2$ (confirmed by MC-simulations, data not shown). However, based on the smaller dimensionless coupling per mode ($\Delta Q/\sqrt{2}$) the respective FC-parabolas are narrower and thus a direct transition into the vibrational ground state from these highly excited states is suppressed. This leads to a slight stabilization of both modes in highly excited states compared to the single mode case.
Additionally, multimode processes allow the redistribution of vibrational energy, i.e.\ the vibrational excitation of a specific mode can be changed by the excitation or deexciation of another mode within a single tunneling event. Consequently, a mode that has left the open-door regime can be brought back into this favorable regime instead of ending up in the vibrational ground state.

Both these mechanisms are even more pronounced in the three mode case as can be seen from the red line in Fig.\ \ref{fig:3m_F_dife}: Due to multimode processes the Fano factor strongly increases with the energy of the electronic level, which is in striking contrast to the one-mode case.
 
Refocussing on Fig.\ \ref{fig:3m_F_09V_05e}, we emphasize again that the two-mode curves representing different pairs of frequencies almost coincide for $\Delta Q \lesssim 4$. This clearly shows that the finding discussed above, that multimode processes increase the Fano factor, does not depend sensitively on the explicit choice of the vibrational frequencies.
For even higher coupling, corresponding to dimensionless shifts $\Delta Q \gtrsim 4$, the behavior
of the Fano factor becomes more complex with no clear dependence on the dimensionality of the problem and the distribution of frequencies, similarly as found above for higher values of $\Delta Q$ in model 1 and model 2.

\subsubsection{Dependence of the Fano Factor on the Relative Electronic-Vibrational Coupling Strength of Two Modes} \label{sec:variation}
\begin{figure}[htbp]
    \includegraphics[width=.6\textwidth]{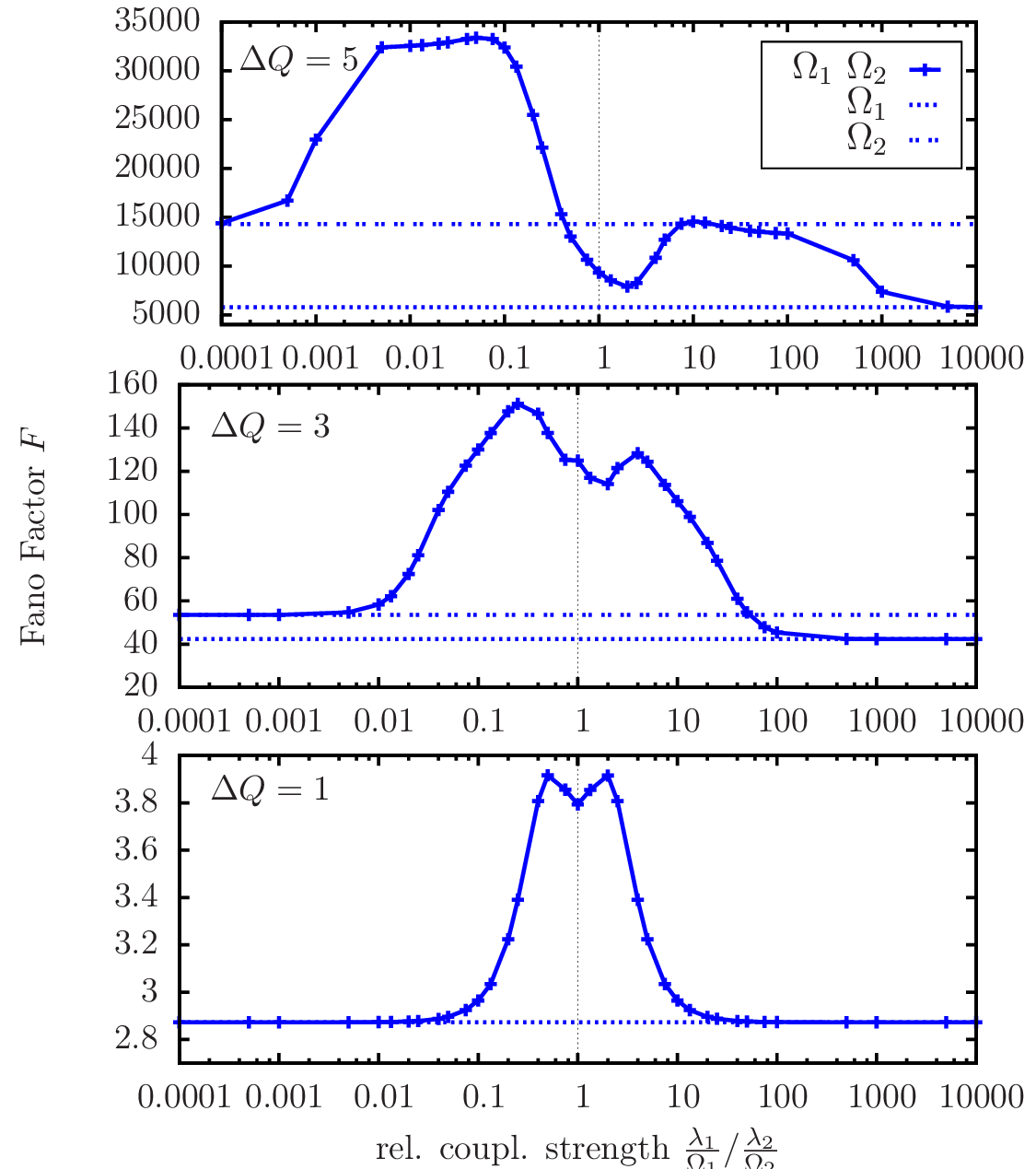}
\caption{Fano factor for model 3 at the three different overall shifts $\Delta Q=1,3$ and 5. Thereby, the relative electronic vibrational coupling strength $(\lambda_{1}/\Omega_{1})/(\lambda_{2}/\Omega_{2})$ is varied, where the dimensionless shift $\Delta Q=\sqrt{ (\lambda_1/\Omega_1)^2 + (\lambda_2/\Omega_2)^2 }$ as a measure for the overall coupling is kept constant. The lines are intended as a guide to the eye.}
\label{fig:2m_difloratio}
\end{figure}

The role of multimode vibrational effects can also be investigated by varying the relative electronic-vibrational coupling strength, but keeping the overall shift $\Delta Q$. For a two-mode model, this means to vary the ratio $(\lambda_{1}/\Omega_{1})/(\lambda_{2}/\Omega_{2})$. To this end, we choose the pair of modes with the frequencies $\Omega_1=85\unit{meV}$ and $\Omega_2=100 \unit{meV}$ and consider the parameters of model 3 discussed above. Fig.\ \ref{fig:2m_difloratio} shows the corresponding Fano for three different constant dimensionless shifts.
In all cases, the Fano factor assumes in the limits $(\lambda_{1}/\Omega_{1})/(\lambda_{2}/\Omega_{2})\rightarrow 0$ and $(\lambda_{1}/\Omega_{1})/(\lambda_{2}/\Omega_{2})\rightarrow\infty$ its one-mode value corresponding to mode 2 or 1, respectively, because only one of the modes is effectively coupled to the electronic state.
It is striking that these limits are only reached for very asymmetric coupling ratios of $10^4 / 10^{-4}$ in the case of $\Delta Q=5$. 
It is also seen that the two limits differ for dimensionless shifts $\Delta Q \geq 3$: The low frequency mode shows the lower Fano factor because principally more excitation / absorption processes can occur than for the higher frequency mode. In the regime of FC-blockade this higher number of processes or the higher net excitation per tunneling cycle, respectively, lead to an earlier termination of the avalanches because highly excited ``terminator'' states ($v \gtrsim 13$) can be reached more easily, as explained in Sec.\ \ref{subsubsec:1m_nonres}.

Analyzing the curves in more detail, it is noticeable that the Fano factor is not maximal for equal coupling to the two modes, but, e.g.\ for $\Delta Q=3$, two peaks can be found for a coupling ratio $(\lambda_{1}/\Omega_{1})/(\lambda_{2}/\Omega_{2})$ of 0.25 and 4. With increasing shift $\Delta Q$ the two local maxima become broader and they are shifted to more asymmetric coupling ratios.
The two peaks can be attributed to the fact that asymmetric coupling to the modes extends the period the system is staying in the open-door stage of an avalanche and thus the avalanche itself:
As already discussed above for symmetric coupling of two modes, multimode vibrational processes allow the more weakly coupled vibrational mode to reach highly exited states in the course of an avalanche where it is stabilized.
In the case of asymmetric coupling this stabilization is even more pronounced because the level of vibrational excitation is comparable with the symmetric case (confirmed by MC-simulations) whereas the dimensionless coupling of the more weakly coupled mode is significantly smaller. Consequently, the probability for direct transitions into the vibrational ground state from these highly excited states is considerably reduced. 

The decrease of the Fano factor for small or large values of the coupling ratio, e.g.\ $(\lambda_1/\Omega_1)/(\lambda_2/\Omega_2) = 0.1$ for $\Delta Q \geq 3$, can be understood in the following way:
In this case the FC-parabola for the more weakly coupled vibrational mode with frequency $\Omega_1$ is very narrow, so that diagonal transitions $v \to v$ are favored. Therefore, this mode cannot be excited to high-lying vibrational states as easily as for, e.g. $(\lambda_1/\Omega_1)/(\lambda_2/\Omega_2) = 0.25$ and thus the stabilization in highly excited states is not as efficient as before. According to our results the stabilization of the more weakly coupled vibrational mode in highly excited states is most efficient for a dimensionless coupling $\lambda / \Omega$ of the order of 0.5, which corresponds to more asymmetric coupling ratios for increasing $\Delta Q$.

In addition, we find for $\Delta Q \geq 3$ that the overall level of the Fano factor, the maximum included, is higher for coupling ratios which are smaller than unity.
This can be attributed to the fact that it is more efficient to stabilize the low-frequency mode in highly excited states as more vibrational processes can occur for this mode and thus even more-highly excited states can be reached than for the high-frequency mode.
This corresponds to the fact that the Fano factor is higher in the limit $(\lambda_1/\Omega_1)/(\lambda_2/\Omega_2) \to 0$ when only the high-frequency mode is coupled.
\subsection{Additional Remarks} \label{sec:remarks}
We close this section with a few remarks. Our results show that, in certain parameter ranges, the shot noise level increases significantly with the number of involved vibrational modes. Although our study is, due to computational limitations, restricted to three vibrational modes, the results suggest that this trend continues for an even larger number of vibrational modes as they are typically found in realistic single-molecule junctions. In this context, it is important to note that the scenario we discuss here involves a discrete set of vibrational modes, which are all similarily and moderately coupled. This situation is completely different from the typical reaction mode szenario, where the coupling to multiple vibrational modes can be expressed by a continous, smooth spectral density and can be translated into a damped reaction mode. In this case, Koch \emph{et al.} \cite{KochPhd,Koch2006} and Haupt \emph{et al.} \cite{Haupt2006} have shown that moderate to strong damping can reduce the shot noise significantly if the phonon bath is thermalized with the electronic leads. However, even in this case,  large shot noise levels can be obtained if the temperature of the phonon bath is significantly higher than the temperature of the leads.\cite{Schaller2013} 

Next, we comment on the range of validity of our results.
In this work, we have discussed avalanche-like transport in the nonresonant as well as in the resonant transport regime. Because the ME approach used neglects higher order effects like cotunneling, we cannot provide a complete picture of all possible processes and effects. Especially, in Sec.\ \ref{subsubsec:1m_nonres}, where we have dicussed the Fano Factor - voltage characteristics at small voltages in the nonresonant regime, our study can only provide an understanding of avalanches based on resonant transport processes. In the deep nonresonant regime, when the chemical potential in the left lead is far below the energy of the electronic state, cotunneling corrections are expected to play a significant role \cite{Thielmann2005,Braggio2006, Aghassi2008} and may lead to a reduction of the shot noise level as shown by Koch \emph{et al.} for a strongly coupled vibrational mode.\cite{Koch2006}
Motivated by experimental results, which report the highest shot noise levels directly before the onset of the resonant transport regime,\cite{Secker2011} we have therefore concentrated on this regime in Sec.\ \ref{sec:results_mm}, where we analyze the coupling to multiple vibrational modes.
In this voltage range, the small molecule-lead coupling chosen, which fulfills the requirement $\Gamma \ll k_B T$, ensures the formal validity of the master equation.
The study of higher order effects on noise properties in molecular junctions, such as, e.g., due to ineleastic cotunneling or cotunneling-assisted sequential tunneling processes, has so far  been restricted to single mode cases.\cite{Thielmann2005,Braggio2006,Leijnse2008,Aghassi2008,Carmi2012,Emary2009} The extension of these more advanced theories to realistic multimode models represents an interesting but also challenging topic for future research.

%
%
%
\section{Conclusions}
In this paper, we have investigated current fluctuations
in vibrationally coupled electron transport through a single-molecule junctions. 
The study considered generic models of molecular junctions comprising a single electronic state
and several vibrational modes and employed a master equation approach developed by Flindt \emph{et al.} \cite{Flindt2005E,Flindt2004}. 
Specifically, we have addressed the physical origin of the large noise levels reported in the experimental studies of Secker \emph{et al.}.\cite{Secker2011}

Earlier theoretical studies of Koch \emph{et al.},\cite{Koch2005,Koch2006} for models with a single 
vibrational mode had predicted that strong electronic-vibrational coupling may give rise to
very high noise levels and the phenomenon of avalanche-like transport. Extending these earlier studies, we have investigated the processes which occur in such  electron avalanches in more detail. Our analysis reveals that two transport regimes can be distinguished: An excitation regime and an open-door regime. In the latter regime, electron transport is enhanced,
because vibrational excitation results in a partial quenching of FC-blockade.
In addition, we have also identified cases, where avalanche-like transport does not reach the open-door regime but is restricted to the vibrational ground and first excited state (``simple'' avalanches).

The major focus of our study was to elucidate the effect of multiple vibrational modes on the noise characteristic of a molecular junction. Our results reveal that the increased complexity of multimode systems  gives rise to novel mechanisms. In particular,  we found two different scenarios where multimode processes result in a significantly larger shot noise level than the coupling to a single vibrational mode: (i) If the electronic level is close to the Fermi level, where in systems with a single vibrational mode only ``simple'' avalanches can occur, multimode processes allow avalanches to reach the open-door stage. Consequently, the avalanches comprise significantly more electrons. 
(ii) If the electronic level is located further away from the Fermi level than typical vibrational energies, avalanches always run in the open-door stage. In this case multimode processes 
drive the system into highly excited vibrational states. A direct transition from these states into the vibrational ground state and thus a termination of the avalanche is suppressed due to the structure of the FC-parabola, which is much narrower than in the one-mode case. Morevover, the vibrational energy can be redistributed among the involved vibrational modes by multimode processes so that a vibrational mode which has left the open-door stage can return into this favorable regime. 

Overall, our results show that in real molecular junctions, which always involve multiple vibrational modes, already weak to moderate electronic-vibrational coupling strength can give rise to high noise levels, especially at the onset of resonant transport as observed by Secker \emph{et al.}.\cite{Secker2011}

\begin{acknowledgments}
This work has been supported by the
German-Israeli Foundation for Scientific Development (GIF)
and the Deutsche Forschungsgemeinschaft (DFG) through
the DFG-Cluster of Excellence `Engineering of Advanced
Materials', SFB 953, and a research grant. RH acknowledges support by the
Alexander von Humboldt Foundation.
The generous allocation of computing time by the computing centers
in Erlangen (RRZE) and Munich (LRZ) is
gratefully acknowledged.
\end{acknowledgments}

\appendix*
\section{Detailed Derivation of the Noise Formula}
%


In this appendix, we give a brief derivation of the noise formula in Eq.\ (\ref{eq:S_0}), following the 
more detailed account in Refs.\ \onlinecite{Flindt2004, Flindt2005E}.
The MacDonald formula \cite{MacDonald1949,Flindt2005E} is given by
\begin{equation}
 S_{I_\tR} (\omega)=  \omega \int_0^\infty \dd t \sin(\omega t) \frac{\dd}{\dd t} [\langle Q_\tR^2 (t)\rangle -\langle Q_\tR (t) \rangle ^2],
\label{eq:app_MDF_S}
\end{equation}
where the regularization
\begin{equation}
  \omega \sin (\omega t) = \lim_{\epsilon \to 0^+} (\omega \sin(\omega t) +\epsilon \cos(\omega t)) \e^{-\epsilon t}
\end{equation}
with the convergence factor $\epsilon \to 0^+$ is implied.
This regularization ensures the correct results for the zero-frequency noise
\begin{equation}
S_{I_\tR} (0)=\lim_{\epsilon \to 0^+} \epsilon \int_0^\infty \dd t  \e^{-\epsilon t} \frac{\dd}{\dd t} [ \langle Q_\tR^2 (t) \rangle -\langle Q_\tR (t) \rangle ^2],
\end{equation}
which is identical to Eq.\ (\ref{eq:MDF_S0}).\\
In order to evaluate the expression for $S_{I_\tR} (\omega)$ in Eq.\ (\ref{eq:app_MDF_S}), we introduce the auxiliary quantity
\begin{equation}
 \tilde S (\omega) = \omega \int_0^\infty \dd t \e^{i (\omega+i \epsilon) t} \left[ \frac{\dd}{\dd t} \langle Q_\tR^2 (t)\rangle -2 \langle Q_\tR (t)\rangle \frac{\dd}{\dd t} \langle Q_\tR (t)\rangle \right],
\label{eq:aux}
\end{equation}
with $\tilde S_{I_\tR} (\omega)= \text{Im} \{\tilde S (\omega) \}= \frac{\tilde S (\omega) + \tilde S (-\omega)}{2 i}$. In a next step, the integration in Eq.\ (\ref{eq:aux}) is interpreted as a Laplace transform evaluated at $s=-i \omega +\epsilon$ and thus we obtain
\begin{equation}
 \tilde S(s)= i s \int_0^\infty \dd t \e^{i s t} \left[ \frac{\dd}{\dd t} \langle Q_\tR^2 (t)\rangle -2 \langle Q_\tR (t)\rangle \frac{\dd}{\dd t} \langle Q_\tR (t)\rangle \right],
\label{eq:aux_s}
\end{equation}
where the regularization by $\epsilon$ is suppressed in the notation.
In order to determine $\tilde S (s)$, we have to evaluate the expressions $\frac{\dd}{\dd t} \langle Q_\tR^2 (t)\rangle$ and $\langle Q_\tR (t)\rangle$. Similarly to the evaluation of the time derivative of the first moment $\frac{\dd}{\dd t} \langle Q_\tR (t)\rangle$ in Eq.\ (\ref{eq:expression1}), the time derivative of the second moment $\frac{\dd}{\dd t} \langle Q_\tR^2 (t)\rangle$ can be related to the current superoperators $\mathcal{I}_\tR^\pm$ and the first moment $\langle Q_\tR (t)\rangle$:
\begin{align}
 \frac{\dd}{\dd t} \langle Q_\tR^2 \rangle&=\sum_n n^2 \dot P_\tR(t,n) \nonumber \\
&=2 \tr{\tS}{\left( \mathcal{I}_\tR^+ - \mathcal{I}_\tR^- \right) \langle Q_\tR (t)\rangle } +\tr{\tS}{\left( \mathcal{I}_\tR^+ +\mathcal{I}_\tR^- \right) \rho^\stat }. \label{eq:expression2}
\end{align}
%

In order to determine the first moment $\langle Q_\tR (t)\rangle$ of the distribution $P_\tR(t,n)$, we define a moment-generating function
\begin{equation}
 F(t,z)=\sum_{n=0}^\infty \rho_n(t) z^n,
\end{equation}
with the properties
\begin{align}
F(t,1)&=\rho(t)=\rho^\sta \label{eq:property1},\\
 \frac{\dd}{\dd z} F(t,z)|_{z=1}&=\sum_n n \rho_n (t)=\langle Q_\tR (t)\rangle \label{eq:property2}.
\end{align}
Substituting Eqs. (\ref{eq:expression1}) and (\ref{eq:expression2}) into Eq.\ (\ref{eq:aux_s}) and subsequently applying Eq.\ (\ref{eq:property2}), the auxiliary quantity $\tilde S (s)$ reads
\begin{equation}
 \begin{split}
\tilde S(s)&= is \left[ 2 \tr{\tS}{\left( \mathcal{I}_\tR^+ -\mathcal{I}_\tR^- \right) \frac{\partial}{\partial z} F(s,z) |_{z=1} } +\langle\langle\tilde 0|\left( \mathcal{I}_\tR^+ +\mathcal{I}_\tR^- \right)|0\rangle\rangle \right.\\
&\left.-2 \tr{\tS}{ \frac{\partial}{\partial z} F(s,z) |_{z=1} } \langle\langle\tilde 0|\left( \mathcal{I}_\tR^+ -\mathcal{I}_\tR^- \right)|0\rangle\rangle \right].
\end{split}
\label{eq:aux_s2}
\end{equation}
Consequently, we are left with the evaluation of the expression $\frac{\partial}{\partial z} F(s,z) |_{z=1}$. To this end, an equation of motion for $F(s,z)$ is derived in Laplace space
\begin{equation}
 [s-\LL-(z-1)\mathcal{I}_\tR^+ - (z^{-1}-1)\mathcal{I}_\tR^-] F(s,z)=F(t=0,z)=\sum_n \rho_n (0) z^n,
\label{eq:FEOM_LT_ord}
\end{equation}
with the initial conditions $\rho_n (0)$. Taking the derivative of Eq.\ (\ref{eq:FEOM_LT_ord}) with respect to $z$ and evaluating it at $z=1$ results in an equation for the quantity $\frac{\partial}{\partial z} F(s,z) |_{z=1}$
\begin{equation}
 (s-\LL) \frac{\partial}{\partial z} F(s,z) |_{z=1} - (\mathcal{I}_\tR^+ - \mathcal{I}_\tR^-) F(s,1)=\sum_n \rho_n (0) z^n.
\label{eq:dF(s,1)}
\end{equation}
In order to determine $ F(s,1)$, we evaluate Eq.\ (\ref{eq:FEOM_LT_ord}) at $z=1$ and thus obtain $ F(s,1)= \mathcal{G}(s) \rho(0)$, where we have used the definition of the resolvent of the Liouvillian $\mathcal{G}(s)=(s-\LL)^{-1}$. After the substitution of $F (s,1)$, Eq.\ (\ref{eq:dF(s,1)}) can be solved for $\frac{\partial}{\partial z} F(s,z) |_{z=1}$
\begin{equation}
 \frac{\partial}{\partial z} F(s,z)|_{z=1}=\mathcal{G} (s) \left( \mathcal{I}_\tR^+ - \mathcal{I}_\tR^- \right) \mathcal{G}(s) \rho(0)+\mathcal{G} (s) \sum_n n \rho_n (0).
\label{eq:diff_F}
\end{equation}
The last step is to evaluate the resolvent $\mathcal{G}(s)$.
To this end, we use the orthogonal projection operator $\PP$, already defined in Eq.\ (\ref{eq:P}), which projects onto the subspace where the Liouvillian $\LL$ is singular.
As the relations $\PP \LL=0=\LL\PP$ and $(1-\PP) \LL (1-\PP)=\LL$ hold, the resolvent $\mathcal{G}(s)$ can be written as
\begin{equation}
 \mathcal{G}(s)=(s-\LL)^{-1}=\frac{1}{s} \PP-\mathcal{R}(s),
\label{eq:resolvent}
\end{equation}
where $\mathcal{R}(s)=(1-\PP) \frac{1}{-s+ \LL} (1-\PP)$ denotes the pseudoinverse in Laplace space.
Substituting Eq.\ (\ref{eq:resolvent}) for $\mathcal{G}(s)$, Eq.\ (\ref{eq:diff_F}) can be written as
\begin{equation}
\frac{\partial}{\partial z} F(s,z)|_{z=1}=\frac{1}{s^2} \PP\left( \mathcal{I}_\tR^+ - \mathcal{I}_\tR^- \right) \ket{0} \rangle +\frac{1}{s} \left[ 1-\mathcal{R}(s)\left( \mathcal{I}_\tR^+ - \mathcal{I}_\tR^- \right) \right] \ket{0} \rangle.
\label{eq:diff_F2}
\end{equation}
Thereby, we have applied the relation $\mathcal{R}(s) \ket{0}\rangle=0$, which directly results from the normalization condition $\langle\langle \tilde 0| 0\rangle\rangle=1$. Additionally we have used the initial condition $\rho_n (0)=\delta_{0n} \rho^{st}$ assuming that we start counting the charge at $t=0$, when the system has already reached its stationary state. Substituting the last expression into Eq.\ (\ref{eq:aux_s2}) and applying the relation $\langle \bra{0} \mathcal{R}(s)=0$, the auxiliary quantity reads
\begin{equation}
 \tilde S (s)=i \left[ \langle\langle\tilde 0|\left( \mathcal{I}_\tR^+ +\mathcal{I}_\tR^- \right)|0\rangle\rangle -2 \langle\langle\tilde 0|\left( \mathcal{I}_\tR^+ -\mathcal{I}_\tR^- \right) \mathcal{R} (s) \left( \mathcal{I}_\tR^+ -\mathcal{I}_\tR^- \right) |0\rangle\rangle \right].
\label{eq:S_s}
\end{equation}
With the relation $s=-i \omega + \epsilon$, $\tilde S(s)$ can be rewritten in terms of $\omega$. Suppressing the regularisation by $\epsilon$ in the notation and taking the imaginary part of Eq.\ (\ref{eq:S_s}), we arrive at
\begin{align}
 S_{I_\tR} (\omega)&=\text{Im} \{\tilde S (\omega) \} \nonumber \\
&=\langle\langle\tilde 0|\left( \mathcal{I}_\tR^+ +\mathcal{I}_\tR^- \right)|0\rangle\rangle-2 \left\{ \langle\langle\tilde 0|\left( \mathcal{I}_\tR^+ -\mathcal{I}_\tR^- \right) \text{Re} \left\{\mathcal{R} (\omega) \right\} \left( \mathcal{I}_\tR^+ -\mathcal{I}_\tR^- \right) |0\rangle\rangle \right\}.
\end{align}
In the zero frequency limit $\mathcal{R} (\omega=0)=(1-\PP) \mathcal{L}^{-1} (1-\PP)$ is real and $S_I(0)=S_{I_\tR}(0)$ holds so that Eq.\ (\ref{eq:S_0}) is obtained.

\bibliography{./mybib.bib}

\end{document}